# A Perspective on Quantum Sensors from Basic Research to Commercial Applications

Eun Oh[1]
Laboratory for Physical Sciences, College Park, MD 20740, USA

Maxwell D. Gregoire[2]
Air Force Research Laboratory, Albuquerque, NM  87117, USA

Adam T. Black[3]
Naval Research Laboratory, Washington D.C. 20375, USA

K. Jeramy Hughes[4]
Fieldline, Inc. Boulder CO, 80301, USA

Paul D. Kunz[5] [6]
Army Research Laboratory, Adelphi MD 20783, USA

Michael Larsen[7]
Northrup Grumman, Woodland Hills, CA 91367, USA

Jean Lautier-Gaud[8]
Exail Quantum Sensors (formerly iXblue), Rue François Mitterrand, Talence, 33400, France

Jongmin Lee[9], Peter D. D. Schwindt[10]
Sandia National Laboratories, Albuquerque, NM 87123, USA

Sara L. Mouradian[11]

1 Physicist, esoh@lps.umd.edu (corresponding author)
2 Senior Physicist, AFRL/RVBYT, Space Vehicles Directorate
3 Research Physicist, Section Head, Quantum Optics, Optical Sciences Division
4 Founder, Chief Executive Officer, Fieldline Industries, Inc.
5 Physicist, Quantum Science & Technology Branch Chief, DEVCOM Army Research Laboratory
6 Department of Physics, The University of Texas at Austin, Austin, Texas 78712, USA
7 Quantum Science Architect, Mission Systems Sector, Emerging Capabilities Development
8 Head of Business Development and Sales, iXblue Quantum Sensors
9 Principal Member of the Technical Staff, Atom-Optic Sensing Department
10 Distinguished Member of the Technical Staff, Atom-Optic Sensing Department
11 Assistant Professor, Electrical and Computer Engineering Department

University of Washington, Seattle, WA, 98105, USA

Frank A. Narducci[12]
Naval Postgraduate School, Monterey, CA 93943, USA

Charles A. Sackett[13]
University of Virginia, Charlottesville, VA 22904. USA

---

12 Professor of Physics, Department of Physics
13 Professor of Physics, Physics Department

Quantum sensors represent a new generation of sensors with improved precision, accuracy, stability, and robustness to environmental effects compared to their classical predecessors. After decades of laboratory development, several types of quantum sensors are now commercially available or are part-way through the commercialization process. This article provides a brief description of the operation of a selection of quantum sensors that employ the principles of atom-light interactions and discusses progress toward packaging those sensors into products. This article covers quantum inertial and gravitational sensors, including gyroscopes, accelerometers, gravimeters, and gravity gradiometers that employ atom interferometry, nuclear magnetic resonance gyroscopes, atomic and spin-defect magnetometers, and Rydberg electric field sensors.

## 1. Introduction to Quantum Sensors

The field of quantum sensing encompasses a wide variety of technological development efforts all with one motivation in common: building sensors with better metrological properties than that which can be achieved classically. Quantum sensors seek to improve upon their classical counterparts in many areas, including precision, accuracy, long-term stability, robustness to environmental noise, and the need for calibration. One of the common aspects of today's quantum sensors is the ability to reference measurements against precisely defined quantities, often ones that are derived in part from unchanging, fundamental constants of nature. Commercialized quantum sensors could provide advanced capabilities in their respective fields such as geophysical exploration [1 - 5] and the healthcare industry [6]; the same is expected of those that have yet to significantly penetrate the market such as quantum devices for navigation [7, 8], space exploration [9], and communication [10, 11].

Commercializing quantum sensors is a challenging endeavor for a variety of reasons. The initial discovery and investigation of the physics principles behind quantum sensors typically starts in a laboratory, often involving expansive optical systems, racks of electronics, and test equipment that consumes kilowatts of power. Initial experiments may rely on expensive, custom-made parts and supporting technologies that range from immature to nearly-non-existent. Devices may need to be hand-assembled by someone with a Ph.D. in physics. The apparatus may only function, at first, in highly-controlled and stable environments. If a quantum sensor is to be commercialized, especially for mobile applications, serious attention must be given to reducing size, weight, power, and production costs while increasing robustness and reliability in harsh conditions. In most cases, reducing such an instrument to something that is portable, rugged, and affordably manufacturable is not a simple matter of shrinking and tightly packing components; doing so requires exploration of new physics.

Because of the difficulty of these problems, significant funding is being invested into this growing industry by both the private and public sectors. Some quantum sensors have already penetrated the commercial market, as shown in Fig. 1, and there are on-going efforts to miniaturize, ruggedize, and commercialize the rest. This article focuses on five sensor types: gyroscopes (Section 3), accelerometers (Section 4), gravimeters (Section 5), magnetometers (Section 6), and electrometers (Section 7).

One of the successful maturations in the quantum sensing industry is quantum magnetometers [12], also called optically-pumped magnetometers, discussed in Section 6. Atomic magnetometers began to see commercial use in the 1960s and 1970s in geophysical [1], archeological [2], astrophysical [9], and defense applications [13, 14]. Today, quantum magnetometers are more compact and, in some cases, able to resolve weaker fields than their classical counterparts [15 - 17, See Fig. 11]. These magnetometers can now be used to conduct magnetoencephalography [3], usually performed by superconducting magnetometers [1, 18 - 20] and therefore requiring orders of magnitude less power and volume because they do not require cryogenic cooling. Quantum magnetometers with the ability to resolve magnetic fields of around 10^-12 Tesla after only 1 second of measurement can now be purchased for under a few thousand dollars [21, 22], supplanting less-sensitive predecessors such as fluxgate magnetometers offered at around the same cost, and are finding uses in communication, navigation, and geological surveys [23].

Gravimeters are another notable success story of quantum-based technology [24 - 29], which are currently being developed and adopted for oil, gas, water, and mineral prospecting [1, 30]. In the last two decades, multiple companies have invested in commercializing gravimeters based on the principles of atom interferometry [30 - 33], demonstrating higher precision and accuracy than the current generation of optomechanical sensors and largely eliminating the need for calibration.

This article will also discuss other quantum sensing technologies that are currently making their way out of the lab and into the commercial world. Quantum gyroscopes and accelerometers [34 – 39] promise real-world applications in aeronautical navigation [7, 8, 40], geophysics [3 - 5], and tests of general relativity [41 - 45]. Quantum electrometers are being investigated for high-precision radio frequency (RF) detection and offer high transparency, wide spectral coverage, and large dynamic range (or range of detectable values of the signals being measured) not achievable with traditional receiver antennas [46].

| Sensor | Architecture | Fundamental research | Laboratory experiment | Portable prototype | Field use | References |
|---|---|---|---|---|---|---|
| Gyroscope | Thermal atom beam interferometer | O | O | O | | 35, 36 |
| | Cold atom interferometer | O | O | | | 135 - 137, 153 |
| | Ultra-cold atom interferometer | O | O | | | 8, 144 |
| | Nuclear magnetic resonance in vapor cell | O | O | O | | 150, 151 |
| Accelerometer | Thermal atom beam interferometer | O | O | O | | 35, 36 |
| | Cold atom interferometer | O | O | | | 137, 153 |
| Gravimeter | Cold atom interferometer | O | O | O | O | 26, 28, 29 |
| Magnetometer | Vapor cell | O | O | O | O | 186, 197 - 200, 205, 209 |
| | Solid-state spin defect | O | O | O | | 215 – 218 |
| Electrometer | Rydberg atoms in vapor cell | O | O | O | | 11, 46, 71, 226, 228, 230 |

Fig. 1  Progress chart showing various stages of development for each quantum sensor to date. Chart data are aggregated from publicly-available commercial information and survey data acquired for this publication.

The first three types of quantum sensors discussed in this article are those that detect inertial and gravitational forces: gyroscopes, accelerometers, and gravimeters. Today, one of the major applications for these types of sensors is inertial navigation [47]. Inertial navigation is a technique in which repeated measurements from gyroscopes (in this case broadly describing devices that measure orientation, rotation rate, and/or changes in rotation rate) and accelerometers (devices that measure velocity changes or gravitational accelerations) over time are used to infer the position and orientation of a vehicle based its known initial position and orientation. Inertial navigation rose to prominence in rocket guidance in World War II [48], but since then found commercial uses in airlines, space flights, and self-driving cars. The advantage of inertial navigation is that it is entirely self-contained and therefore can be employed at any time regardless of external conditions such as visibility or limited access to radio signals such as GPS signals.

Gravimeters can be thought of as single-axis accelerometers optimized to precisely measure gravitational acceleration in a geospatial location. In gravity surveying applications, the most important characteristics of a gravimeter are absolute accuracy and the ability to maintain that accuracy over long durations (i.e., low long-term drift). Because gravimeters usually do not require higher measurement rates or need to be able to operate in any geometric orientation, scientists and engineers are able to optimize the precision and accuracy of these devices by restricting their design such that they can only operate effectively in a "vertical" orientation and can typically only measure accelerations that are approximately equal to 1 $g$ (see Section 5 for further explanation).

Gravimeters can also operate in different locations in tandem to measure gravity gradients, or changes in the force of gravity over an area [43, 49, 50]. In this case, these devices also benefit from common-mode noise rejection in order to achieve even higher precision and accuracy. Gravity gradients carry additional information about the mass distribution in an area of interest beyond what can be communicated by a single measurement. In addition to discovering subterranean regions of water, oil, gas, and minerals, high-precision gravimeters and gravity gradiometers can be used to detect underground voids and tunnels [51]. Gravity gradiometry can also be useful in navigation. Gravity gradiometers on an aircraft or spacecraft can be used to distinguish between gravitational and inertial acceleration–something that is forbidden by the equivalence principle [52] for single measurements–and can be used to implement gravity-aided navigation when combined with a map of Earth's gravity.

For all of the aforementioned applications, scientists and engineers seek to improve precision and accuracy in the next generation of inertial sensors. *Precision*, also known as *statistical uncertainty*, describes the extent to which noise in the sensor, such as the random motion of electrons in a photodiode or random jitters in the wavelength of light produced by a laser, leads to similarly random deviations, or *statistical errors*, in a measurement. Precision can also be negatively affected by external noise sources such as vibrations or, if considered over the course of minutes or hours, temperature changes. High precision implies that a single measurement of a quantity can be made with the assurance that only very small, random (or pseudo-random) statistical errors might have perturbed the measurement, and that it would only take a relatively short time to average a series of measurements down to some target resolution. Precision is often characterized by *sensitivity*, which describes how well a measured quantity can be resolved as a function of the time over which repeated measurements of that quantity are averaged. If statistical errors in a measurement can be approximated as being sampled from a Gaussian distribution, the precision of a measurement improves as the square root of the averaging time. The proportionality factor relating measurement precision to the square root of the averaging time is the sensitivity.

*Accuracy*, also known as *systematic uncertainty*, describes the extent to which a sensor's measurement of a quantity–after averaging out statistical errors–systematically deviates from the true value of that quantity. Higher accuracy implies that a device's measurement (again, after averaging out statistical fluctuations) of a particular quantity, such as gravity, is closer to being objectively correct. Another aspect of systematic uncertainty is *stability,* or how systematic errors may change over time. Repeated measurements of a constant quantity might change over long periods due to changes in nominally-constant parameters internal to the measuring device, such as the temperature of a photodiode or the physical distance between optical elements. If a device with high stability is used to measure the same quantity multiple times over a long duration, the measured value will change very little. *Stability* is typically broken down into quantities called *bias stability*, which describes the stability of long-term measurements of a null value, and *scale factor stability*, which describes drift in the proportionality constant between the reported measurement and the quantity being measured. Note that it is possible to have high stability, or highly stable systematic uncertainty, even if that systematic uncertainty itself is high. Many quantum sensors have the potential to derive both high accuracy and high stability by referencing measurements to unchanging, well-known constants of nature, such as the properties of certain atomic species.

Precision is especially important in inertial navigation. When acceleration and rotation data are repeatedly acquired and integrated to infer the motion of a vehicle, tiny statistical errors in the measurements can add up quickly, causing even a highly-accurate inertial navigation system to accumulate significant position errors if used for a long enough time [47]. Similarly, high precision does not necessarily guarantee high accuracy. Inertial sensors, especially gravimeters, are often described as making "relative" or "absolute" measurements. A "relative" measuring device, even if very precise, is assumed to have both fixed and time-varying *systematic errors* in its measurements. Such a device may converge on a measurement very quickly due to having little measurement noise, but that measurement will likely have large systematic uncertainties, which must be *calibrated* via comparison with a more accurate device. Conversely, an "absolute" measuring device is highly accurate and is assumed to require no calibration.

Quantum inertial sensors are currently in various stages of technological maturity. Nuclear magnetic resonance (NMR) gyroscopes [53, 54] currently exist as centimeter-scale, ruggedized units that could be integrated into a larger system, though development of the product is ongoing. Atom interferometer gravimeters are productized into larger devices suitable for an aircraft or truck and used to make real measurements of the variations in the density of Earth's crust. The oil and gas industries are interested in additional work to make these systems more mobile. There are a few different designs of gyroscopes and accelerometers based on atom interferometry, each achieving different precisions and bandwidths depending on how the atoms are handled within the device. The most mature versions, based on interferometry with thermal atom beams (discussed in Section 3.1), are actively being developed in the commercial world for future space flight tests [55].

As mentioned earlier in this article, another major category of quantum sensing is magnetometry, or the detection and measurement of magnetic fields. In Section 6, we focus on atomic vapor-cell magnetometers [12] and *spin-defect magnetometers* [56 - 58], the latter of which behave similarly to atomic systems. Vapor-cell-based atomic magnetometers, which measure the effect that magnetic fields have on the states of atoms in hot gasses, have made tremendous progress in the recent decade due to investment from the biomedical industry [59]. They can measure magnetic fields smaller than a $10^{-15}$ Tesla in a shielded environment [12, 16, 17], or roughly 10 billion times weaker than Earth's magnetic field. Spin-defect magnetometers are operated similarly to vapor cell magnetometers: instead of measuring the effects of magnetic fields on atomic states, spin-defect magnetometers measure the effects of magnetic fields on the atomic states of defects, or impurities, in a crystal lattice. While these devices continually develop towards high precision [60], their solid-state design and room-temperature operability make them very attractive. These lab prototypes promise even further reduction in size and complexity and may enable micro-scale magnetometry for materials, circuit, and biological analysis. Spin-defect magnetic field sensors have been demonstrated in Sapphire [61] and Silicon Carbide [62], but the most technologically advanced solid-state quantum magnetometers exploit the molecular structure of nitrogen-vacancy (NV) centers caused by nitrogen impurities in diamond [63]. These devices have been shown to provide unrivaled spatial resolution and can be used to map the magnetic fields at microscopic levels.

In addition to biomedical applications, highly-sensitive quantum magnetometers may enable communication, navigation, and situational awareness. Magnetometers have the ability to detect the magnetic components of radio signals (which are composed of electric and magnetic fields) over a very wide range of frequencies. Radio signal detection capability is especially important when the user needs to detect very long-wavelength signals, which can require antennas ranging in size from meters to kilometers. For example, a fist-sized magnetometer acting as a radio receiver can replace a very low frequency (VLF) antenna apparatus weighing thousands of kilograms and stretching out over a kilometer while maintaining the same environmental-noise-limited signal-to-noise ratio [64].

Taking advantage of high bandwidth, quantum magnetometers may also aid in navigation systems. Today's magnetometers are sensitive enough to measure small deviations in the strength of Earth's magnetic field. If those measured deviations are compared to existing maps of Earth's magnetic field, a navigating vehicle can ascertain its position without access to external signals, landmarks, or GPS satellites. Finally, portable, precise magnetometers could be used to detect nearby hidden metal objects such as contraband items at a border checkpoint, shovels in an underground tunnel, or a submerged submarine [65, 66].

Complementing magnetic field detection, a similar type of quantum sensor is used to detect electric fields. Quantum electrometers, discussed in Section 7, detect electric fields by observing changes to the atomic energy levels of highly-excited atoms called Rydberg atoms [67, 68]. The energies of Rydberg states can be interrogated to sense electric fields or the electric component of radio signals anywhere in an incredibly wide range of frequencies, from DC to hundreds of Gigahertz [46, 69]. Rydberg technology can be used to scan for communication signals over a very wide band and with high dynamic range (i.e., the ability to measure both very weak and very strong signals without modifications to the device) [70, 71]. As with quantum magnetometers, Rydberg based quantum electrometers could be used to detect low frequencies that would otherwise require large sized classical antennas to detect [72]. Additionally, these devices can be used to detect frequencies much higher than what is easy to achieve with conventional electronics, potentially simplifying and miniaturizing future communication systems that rely on frequencies above around 100 GHz [69]. Rydberg electrometry has seen steady development in academic laboratories over the past few decades, gaining traction in the commercial world in the past handful of years.

In this paper, we give an overview of these aforementioned sensors in terms of their commercial applications so that readers can be aware of their status and recent progress. For each technology, we present motivations and use cases, technological principles, and associated challenges. Finally, in the last section, we summarize how engagement between scientists and engineers working on quantum sensing technologies and members of the aerospace industry could speed development and ensure that the aerospace industry's needs are being addressed. The authors would like to note that the scope of this paper is limited to quantum sensors based on the manipulation of atomic or atom-like states that are currently in the process of being transformed into commercial products. We do not cover current academic research into such sensors, which is deserving of its own review. Also, we note that the scope of our paper is limited to sensors exploiting the aforementioned physics and does not include other sensors that rely on other quantum effects such as quantum Hall effect devices [73], superconducting quantum interference devices (SQUIDs) [1, 19, 72], or quantum-optical gyroscopes [74], to name some examples. Atomic clocks [75], which rely on the physical principles discussed in this article, will also not be discussed.

## 2. The Physical Principles of Atom-Light Interaction and Atom Interferometry

Before delving into discussions of the quantum sensors themselves, it is necessary to understand some of the underlying physics that is common to multiple sensor architectures. The foundation of all of the quantum sensors discussed in this article is atom-light interaction [76]. In these sensors, specific frequencies of light are used to measure or exploit the properties of atoms and molecules. One of the foundational discoveries in the early 1900s that led to today's understanding of quantum mechanics was the realization that the energies with which electrons orbit atoms – determined primarily by the kinetic energy of those electrons and the electrostatic potential energy between electrons and the nucleus – are *quantized* [77]. This means that those atomic energy levels are restricted to specific values. This realization helped explain a previous discovery in atomic physics, which was that atoms appeared to absorb and emit only certain wavelengths of light [78]. Further study led to the understanding that, for an atom's energy to increase from a lower level to a higher level, it must absorb a photon with energy equivalent to the difference between those energy levels. Similarly, an atom in a higher energy level can decrease its energy to a lower level by emitting a photon equal to the energy difference between levels. That emission can occur spontaneously (i.e., at a random time and in a random direction), or it can be stimulated by interaction between the atom and another photon of similar energy [79]. Stimulated emission is the key physical principle that enables laser technology [80]; when a photon stimulates an atom to emit another photon, that second photon will have the same frequency, phase, and direction of propagation as the first. In a laser, this effect is used to create highly-directional beams composed of large numbers of photons with identical frequency and phase.

A straightforward way to observe atom-light interaction is with atoms in a *vapor cell*, which is a small (few cubic centimeter), transparent chamber with a dilute atomic gas. Light of specific wavelengths from lasers or lamps can be directed into the vapor cell to interact with the atoms. To measure atomic energy levels or changes in those levels, one can either measure the extent to which the incident light is absorbed or observe light that is scattered or emitted by the atoms in the cell. Using this basic framework, anything that would change the atomic energy levels can be detected and measured. For example, atomic energy levels may change by specific amounts in the presence of external electric [81, 82] or magnetic fields [83]. The frequencies of light that interact with an atom can also be influenced by the orientation of that atom as it rotates with respect to the apparatus. The following sections will discuss how these effects can be used to measure electric and magnetic fields as well as rotation rates [53]. Vapor cell technology was the basis for many early atomic clocks [75], and commercialized atomic clocks still rely on these techniques. Vapor cells are conducive to making high-precision measurements because all of the atoms of a given isotope drifting around in the cell are truly identical. If all of those atoms are exposed to an identical external field, they will all react in the same way.

Quantum sensors have the potential to improve upon their classical sensors in terms of accuracy and precision. Quantum sensors that rely on atom-light interaction achieve high accuracy through the homogeneity of atomic properties, i.e., the fact that every atom of the same isotope is identical and indistinguishable. Following the laws of quantum mechanics, the energies of the electrons in their various orbitals around the nucleus and the required energies to excite those electrons using optical, magnetic, and electric fields are well defined. For example, shining a specific wavelength of light on different atoms of a single isotope will always have the same probability to excite electrons from one specific state to another over a wide range of conditions. This homogeneity leads to a high degree of stability in sensors that employ these techniques, meaning that there will be relatively little drift in repeated measurements of a constant quantity over a long period of time. Note that drift can and does still occur in practical applications because, for example, completely shielding from all stray electric and magnetic fields can be a difficult task.

High precision is typically derived from a system having a strong response to changes in the quantity being measured or from being able to measure that response very precisely. Precision depends on the specifics of the quantity being measured and the method of doing so. For example, the index of refraction of a gas of highly-excited atoms (called Rydberg atoms, discussed in Section 7) in a vapor cell can change by 10^15 times more than that of a solid material. This effect gives vapor-cell-based, quantum electric field sensors a huge advantage in sensitivity that, in certain frequency ranges, allows them to compete with classical antennas that are significantly larger.

Atom-light interaction can be the basis for more complex measurement techniques as well. The first three sensor types presented in this article–gyroscopes, accelerometers, and gravimeters–can all be implemented using atom interferometry [84, 85]. To understand how atom interferometers have the potential to improve upon the precision of their classical counterparts, it helps to understand the benefits of interferometry in general.

Consider a beam of light freely propagating through an apparatus. If the apparatus undergoes an acceleration or rotation, the waves of light may appear to be deflected with respect to the apparatus. In a non-interferometric deflection measurement, the precision with which that deflection can be measured depends largely on uncertainties in the dimensions and positions of apertures and beams in the apparatus and is typically limited to within a few hundred nanometers at best [85].

In an interferometer, the propagating waves are split and recombined. Where they overlap upon recombination, a periodic arrangement of bright and dark spots called an *interference pattern* forms based on the extent to which the waves interfere *constructively* (the waves add together to form a wave with larger amplitude) or *destructively* (the waves subtract from each other to form a wave with smaller amplitude). If the interfering waves shift to interfere more constructively or more destructively, the position of their interference pattern appears to change. Fig. 2 Illustrates how any change in position of that periodic interference pattern in a Mach-Zehnder interferometer corresponds to the change in the classical trajectory of the propagating waves. In the top diagram in Fig. 2, rotation of the apparatus causes deviations in the trajectories of the split beam as it passes through the interferometer: rotations cause a Coriolis force to affect trajectories, while accelerations (not pictured) cause parabolic deviations. Those deviations cause the interference pattern to be displaced in space by a corresponding amount. In the bottom diagram, rotation of the apparatus changes the effective path length of each leg in the interferometer, thereby changing the spacing between wave fronts (indicated by the thin, repeated lines perpendicular to the wave trajectories) with respect to the apparatus. The key advantage is that the deflection of the interference pattern can be measured as a small fraction of its spatial period, typically expressed as a *phase shift*. Since the spatial period of an optical interference pattern may be as small as the optical wavelength (hundreds of nanometers to a few microns in length), the deflection of that interference pattern can be measured to distances orders of magnitude smaller than a single wavelength of light–and therefore orders of magnitude smaller than what can be achieved classically.

Interferometry can be accomplished with atoms as well. The de Broglie hypothesis [86], also referred to as wave-particle duality in quantum mechanics, implies that atom waves (also called matter waves or atomic wave packets) should also be able to diffract and form interference patterns when their amplitude or phase is affected, where the intensity or "brightness" of a particular point in the pattern is described by the number density of atoms present. In one of the most common types of atom interferometers, the light-pulse atom interferometer [87, 88], depicted in Fig. 4, atom waves are split, redirected, and recombined to construct matter wave interference patterns using a tailored light pulse sequence that interact with the atoms in specific ways. These light pulses address specific atomic internal states and change the atoms' momenta depending on those states. Their frequencies and durations can be tuned to give the atoms a desired probability to transition from one state to another [89]. For example, a light pulse can be used to effectively split an atom wave by placing the atom in a *quantum superposition* of two momentum states, each corresponding to a different internal state of the atom and a different trajectory [90]. Put more intuitively, atoms in a quantum superposition of two states–in this case, states that represent different momenta–are said to have some probability of being in each state. Only when those superposition states are *observed*, or more generally, made to interact with a part of their environment such as a detector, are the atoms said to exist in one state or another. In the light pulse sequence of a typical atom interferometer, atom waves are first split by a light pulse into diverging trajectories. Then those trajectories are reversed with a second pulse so that they are made to re-converge at a point in space. Finally, the waves are recombined with a third pulse identical to the first pulse. After the third laser pulse, the fraction of atoms in each momentum state contains information about the phase difference between the two trajectories, which in turn contains information about the inertial or gravitational forces experienced by the atoms. After a period of time, the atoms in different momentum states will become spatially separated from one another. Measurement is typically accomplished by a technique called *absorption imaging*, which requires a camera to image atoms that have absorbed light from a laser beam resonant with their atomic transitions.

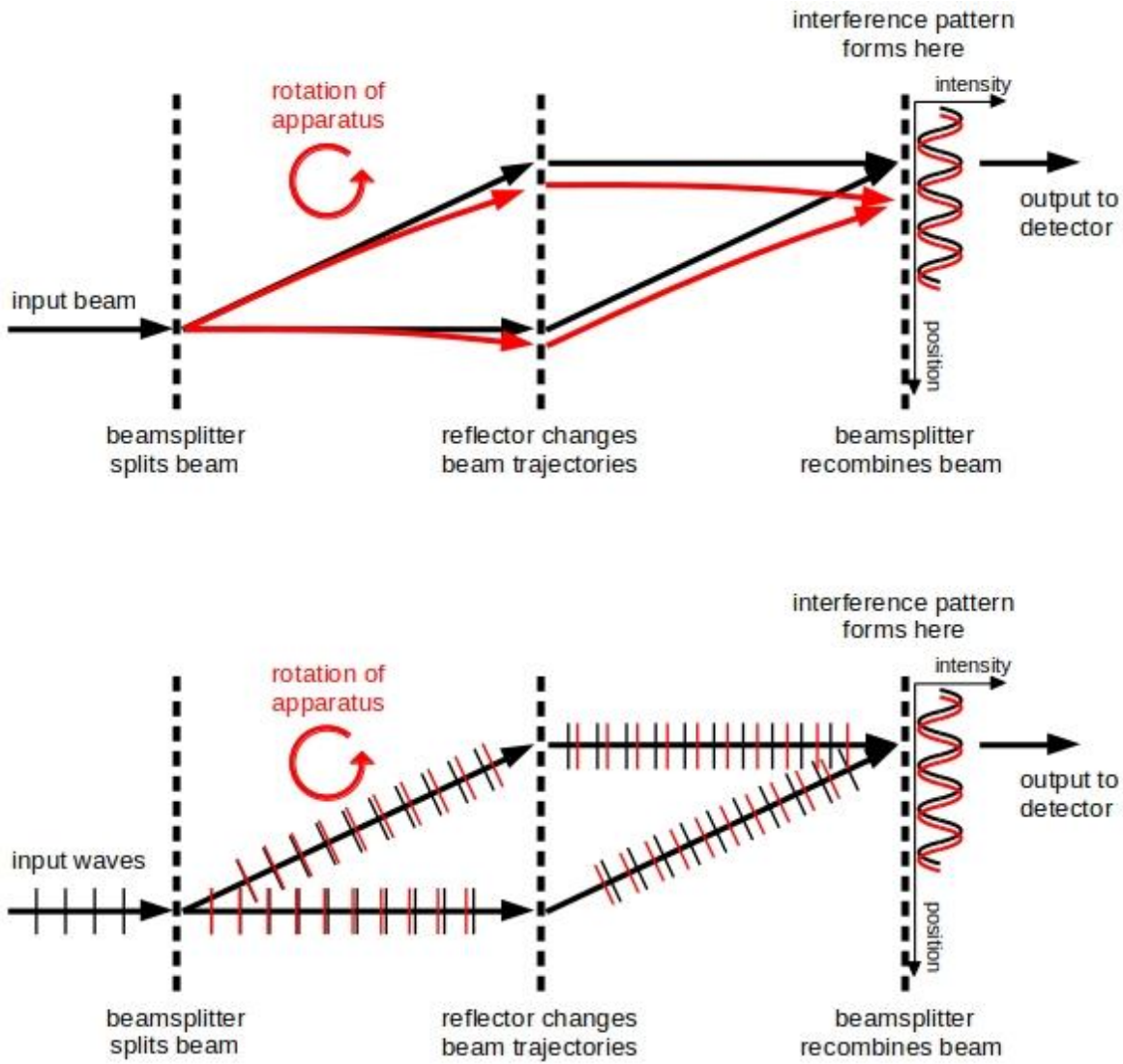

Fig. 2 Two different ways to visualize the effect of rotation on a Mach-Zehnder matter-wave interferometer.

One advantage of an atom interferometer gyroscope over an optical interferometer gyroscope with the same fixed configuration of split and recombined beams is that atoms have several orders of magnitude lower velocity than light waves. This allows for a longer *interrogation time*, or the amount of time between when the atom waves are split and when they are interfered during which they are exposed to inertial forces. As interrogation time increases, the trajectories of those atom waves are deflected by larger distances under the influence of inertial or gravitational forces, increasing sensitivity to those forces. Said another way, the longer interrogation time results in a larger phase difference between the two trajectories. This is one of the main reasons that atom interferometer gyroscopes have the potential to outperform classical sensors.

One limitation on the precision of an atom interferometer is the temperature of the atoms during the interferometry process [91]. In an ideal scenario, all the atoms passing through the interferometer would have exactly the same starting trajectory and therefore contribute to the formation of the exact same interference pattern. In reality, however, the atoms will have some distribution of initial velocities, which will result in different atoms contributing to the formation of a distribution of different interference patterns, each with slightly different length and orientation within the apparatus. Having a multitude of different, overlapping interference patterns will have the effect of "washing out" the signal, adding noise to the measurement [92]. Although the washout effect can be mitigated by reducing the interrogation time in the interferometer, doing so results in a corresponding decrease in precision. The solution to this problem is to reduce the velocity spread of the atoms or, equivalently, to reduce their temperature. There are several ways to do so, each with their advantages and disadvantages.

The simplest way to implement an atom interferometer is to tightly collimate a beam of atoms. This technique is known as *thermal beam* atom interferometry [93] because the atom beams are typically produced by heating atoms to a vapor inside a small oven and spraying them out of a small hole at velocities of hundreds or thousands of meters per second. The temperatures of the atoms in the directions perpendicular to the beamline can be reduced by passing the beam through one or more narrow apertures. Doing so has the effect of narrowing the transverse velocity spread of the atoms (or the distribution of atom velocities in directions perpendicular to the beam), even while the velocity spread of atoms in the direction of the beam remains relatively wide. In the relatively extreme case of a 2-meter-long atom interferometer, the temperature of the atoms' motion along the beamline can be thousands of degrees, but due to the tight collimation of the high-velocity beam, the temperature of their transverse motion can be as low as a few tens of microKelvin [94] (for reference, room temperature is around 295 Kelvin).

Another way to achieve low transverse temperatures in a thermal beam is to implement a form of two-dimensional laser cooling. A laser-cooling technique known as *optical molasses*, which makes use of physical mechanisms such as *polarization-gradient cooling* or *Sisyphus cooling* [95, 96], can be applied to the atoms in the thermal beam in order to reduce their transverse temperatures beyond what could be reasonably achieved with collimation in a small apparatus. A pair of counter-propagating lasers–in this case propagating perpendicular to the atom beam–with specific polarizations is used to create a dynamic optical field that opposes the motion of atoms in that field if they travel in either direction along the beams. In a different 2-meter-long atom interferometer, optical molasses was used to achieve transverse velocity spreads as low as 10 cm/s, corresponding to temperatures on the order of 100s of microKelvin [35]. While this transverse temperature is an order of magnitude larger than in the previously discussed 2-meter-long interferometer in reference [94], it allowed the researchers to achieve reasonably low temperatures in conjunction with high atom fluxes of 10^8 atoms/s versus only 10^5 atoms/s in reference [94].

Following collimation or cooling, atom interferometry can then be performed in the transverse direction, taking advantage of relatively low transverse temperatures and minimizing exposure to the high longitudinal temperatures that result from the oven. This method was employed in the first atom interferometers used to make high-precision measurements of atomic properties [97]. In addition to their simplicity and ease of manufacturing, thermal beam atom interferometers also have the advantage of producing a quasi-continuous stream of data due to the quasi-continuous stream of atoms flowing through the system. As a result, they can achieve high bandwidths of multiple hundreds of Hz, which is to say that they can detect relatively fast changes in the forces they measure.

Another way to achieve low temperatures is with a combination of *magneto-optical trapping* (MOT) [98] and optical molasses. In this article, we refer to devices that use optical molasses as their final form of cooling as *cold atom* devices. Invented in 1987 and eventually leading to a Nobel prize, a MOT is capable of producing atoms that are cooled in all three dimensions to around 100 X 10^-6 Kelvin. The technique is an extension of Doppler cooling [99], in which atoms absorb lower-energy laser light and re-emit higher-energy light. The energy deficit between absorption and re-emission is unbalanced using the thermal and kinetic energy of the atoms, which results in the atoms being cooled. In a MOT, the addition of magnetic fields and the use of specific laser polarizations creates a situation in which atoms become trapped as they are being Doppler cooled. Atoms experience a spatially-dependent force that pushes them towards a local minimum in the magnetic field that defines the center of the trap. MOTs can produce cold beams of atoms or cold clouds that can then be released into freefall or launched to propagate in specific directions. After cooling in a MOT, atoms are further cooled down to temperatures on the order of 10^-6 Kelvin–below the Doppler cooling limit–via optical molasses.

One of the key differences between MOTs and thermal beams lies in the trade-off between performance and complexity. Thermal beam atom interferometry has the advantage of simplicity over MOTs, which require complex, high-power, multi-frequency laser systems. On the other hand, MOTs can achieve lower temperatures than the transverse temperatures in collimated, thermal beams, and those low temperatures can also be realized in all three dimensions. MOT-based sensors also often have limited bandwidth compared to thermal-beam-based sensors. Many MOT-based sensors rely on shot-by-shot measurements in which a cold cloud is produced, released, interfered, and measured before the cycle repeats [100]. With efficient atom recapture processes in place, bandwidths as high as a few hundred Hz [38, 100] can be reached. In this case, the bandwidth of the sensor is limited by the time required to produce a cold cloud as well as the interrogation time of the interferometer. In Sections 3 and 4, methods for reducing that cycle time are discussed.

Finally, the smallest atomic velocity spreads are available in Bose-Einstein condensates [101] or so-called "ultra-cold" atom clouds. A Bose-Einstein condensate (BEC) is an ensemble of very low-temperature, low-density atoms. At temperatures near absolute zero, these atoms occupy the lowest quantum state *en masse*, exhibiting macroscopic quantum properties such as interference. BECs are formed using methods that cool atoms to temperatures lower than what is achievable with optical molasses, often to the ground state of an optical or magnetic trap [102]. The process of Bose-Einstein condensation usually begins with magneto-optical trapping and may end with an RF-induced process called evaporative cooling [103] or a laser-cooling process called Raman sideband cooling [104], though other viable processes exist. If a BEC is not specifically needed in a given apparatus, these cooling methods can be employed in order to cool atoms to temperatures referred to as "ultra-cold." While the temperature of a pure BEC cannot be defined, the temperatures of ultra-cold atoms acquired using a BEC production process is less than a few hundred nanoKelvin. BEC-based systems take the trade-off of sensitivity vs complexity and bandwidth to an extreme level. While ultra-cold temperatures have the potential to lead to the highest-precision measurements, BECs take additional infrastructure on top of what is required to produce a MOT, which adds additional size, weight, power, and complexity. The time required to produce a single BEC in systems advancing toward commercialized status is on the order of seconds [105], which makes for extremely slow sensing measurement cycles. However, recent research has demonstrated continuous BEC production [106].

In the sections below, the authors discuss technology gaps that are currently inhibiting the size-reduction, power-reduction, and ruggedization of quantum sensors. The foremost contributor is the lasers. The quantum sensors discussed in this article operate by manipulating atomic or molecular energy states with laser light. Unfortunately, the set of laser wavelengths required to interact with those atomic or molecular states do not necessarily overlap with the set of laser wavelengths that are convenient to produce. However, recent advances in photonic integrated circuits and semiconductor photonic devices [107 - 110] will likely streamline the manufacture of quantum sensors and enable further reductions in size. Another key supporting technology is vacuum technology. The atom interferometers discussed in this article utilize vacuum chambers in order to maintain low atom temperatures and limit decoherence [111]. Unfortunately, vacuum pumps tend to be fairly large and heavy, consume significant amounts of power, and produce stray electromagnetic fields that modify atomic states in unwanted ways. Fortunately, improved vacuum chamber and vacuum pumping technology is becoming available, including passive pumping technologies and low-helium-permeability materials [112 - 114].

It is important to note that, despite the use of the words "cold" and "ultra-cold" and the discussions of atoms with extremely low temperatures, none of the aforementioned cooling techniques involve cryogenics, or the refrigeration of bulk material or components. While collimating apertures, light, and magnetic fields can be used to remove kinetic energy from the atoms themselves, the apparatus itself remains at the temperature of its surroundings.

## 3. Atomic Gyroscopes

Currently, the most precise gyroscopes available commercially are ring laser gyroscopes (RLGs), fiber-optic gyroscopes (FOGs), and various mechanical gyroscopes [115, 116].  RLGs and FOGs are based on the principles of optical interferometry. These mature technologies have received decades of innovation support and are currently employed in space, industrial, defense, and consumer applications. In this article, we discuss two quantum approaches challenging the precision and accuracy of optical gyroscopes: atom interferometry and nuclear magnetic resonance (NMR). Atom interferometers track the rotation-induced deviations in the trajectories of atoms propagating through space. NMR gyroscopes observe the precession of spinning atomic nuclei about magnetic field lines caused by rotation of that magnetic field and, by extension, the apparatus that produces it. A comparison of various gyroscope technologies is shown in Fig. 3. The red labels in Fig. 3 denote two efforts by DARPA (Chip-Scale Combinatorial Atomic Navigator (C-SCAN) [117] and Precision Inertial Navigation Systems (PINS) [118]) to achieve certain combinations of precision and volume. Those red markers only indicate program objectives, not descriptions of actual devices.

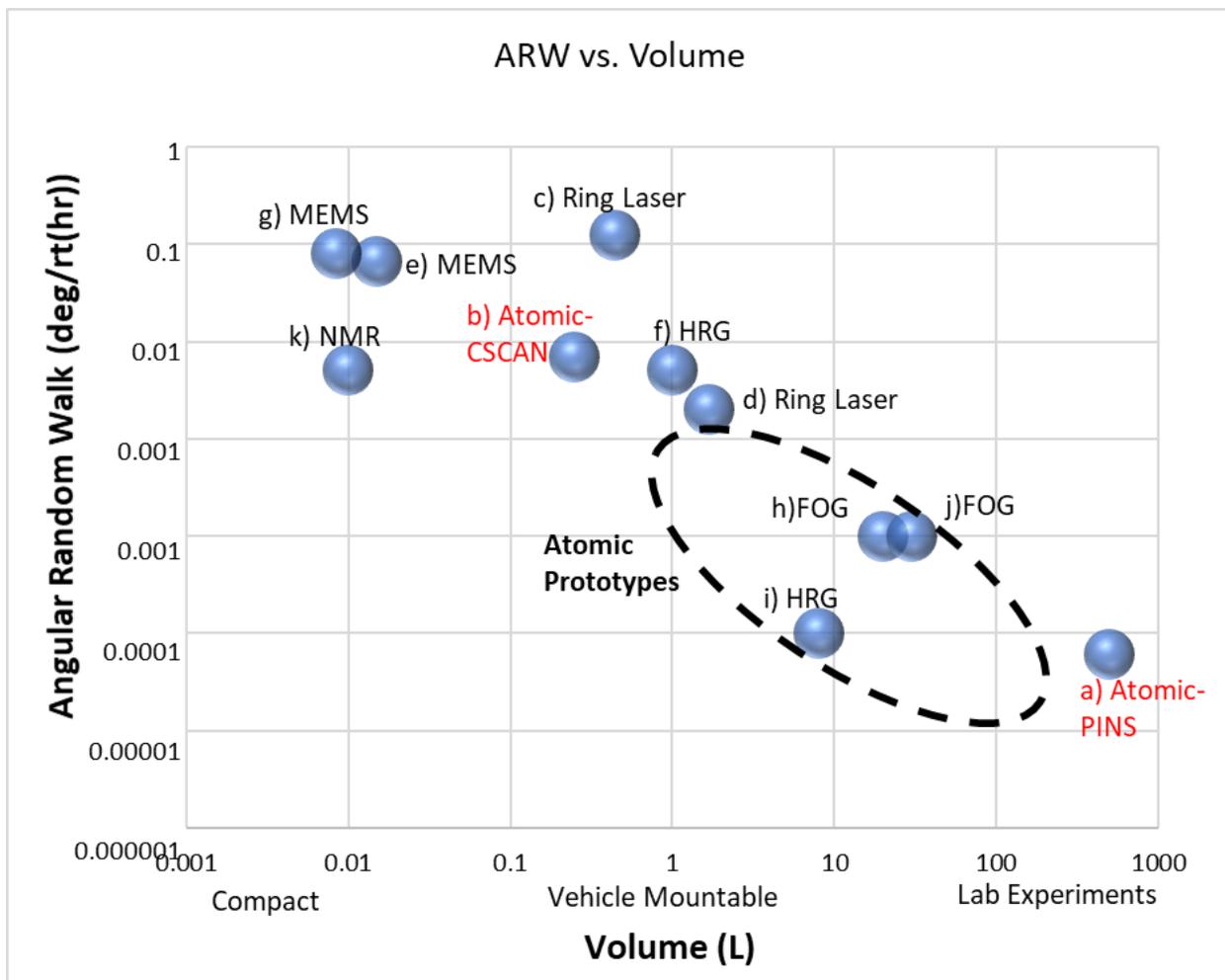

Fig. 3  Gyroscope precision, characterized as *angle random walk*, vs full-system volume for both commercially available gyroscopes and quantum gyroscopes currently being commercialized.

References: a) [117], b) [118], c) [119], d) [120], e) [121], f) [122], g) [123], h) [124], i) [125], j) [126], k) [8].

Commercialized, all-optical gyroscopes as well as atom interferometer gyroscopes both operate based on the Sagnac effect. The Sagnac effect describes how rotation of an interferometer shifts the phase of its interference pattern. Fig. 2 illustrates this effect. If the apparatus is rotated about any axis in the plane of the area enclosed by the two paths, the waves in one path will traverse a longer distance than the waves in the other path in the reference frame of the apparatus. Equivalently, the classical trajectories of those paths as well as the interference pattern they form upon recombination are deflected with respect to the apparatus as the latter rotates with respect to the former. In either picture, the rotation causes a shift of the interference pattern's phase that is proportional to the enclosed area and the rotation rate and inversely proportional to the velocity and wavelengths of the interfering waves. In an optical interferometer, the speed of light is very high (around 300 million meters per second) and the wavelength of the light is close to $10^{-6}$ m. In an atom interferometer, the speed of the atoms can be anywhere from 1000s of meters per second to less than 1 meter per second, implying that atom interferometry carries the advantage of longer interrogation time, as discussed in Section 2. Similarly, the wavelengths of the quantum mechanical atom waves can be less than $10^{-12}$ m in length. This gives an atom-based Sagnac interferometer a phase shift on the order of $10^{11}$ larger than a light-based one measuring the same rotation rate. This improvement in *scale factor*, or measurable response to the quantity being measured, is one of the main motivations for developing atom interferometer gyroscopes because it implies that atom interferometer gyroscopes should be able to achieve much higher precision than their optical predecessors.

However, it is important to note that the apparent multiple-order-of-magnitude advantage in scale factor does not tell the whole story. First, the flux of atoms in atom interferometers is several orders of magnitude less than the typical photon flux in optical interferometers. So even if an optical interferometer were to experience a much smaller interference pattern phase shift than a comparable atom interferometer, the higher photon flux could still result in higher resolution of that phase shift and, therefore, a higher-precision determination of rotation rate. Second, achieving large enclosed areas to exploit the Sagnac effect is typically much more difficult with atoms than it is with light, whereas the area in a ring laser gyroscope can be made very large [115] and the fiber in a fiber optic gyroscope can be wound into many loops with a very large total enclosed area.

The large enclosed area of optical gyroscopes, however, has its own drawbacks. The long-term accuracy of optical gyroscopes is strongly dependent on the stability of the index of refraction of the medium in which the laser propagates. Environmental effects such as temperature changes or ionizing radiation can change that index of refraction, changing the speed of light through the interferometer and limiting the stability of an optical gyroscope. Fiber optic gyroscopes with multiple loops of fiber optic cables suffer especially from this effect. Contrarily, atom interferometer gyroscopes show excellent long-term accuracy. Atom waves in an interferometer propagate through vacuum and thus do not contend with index changes. Furthermore, the trajectories of atom waves are largely determined by photon momenta that are tied to fixed, atomic resonances.

Gyroscopes based on ultra-cold atoms or Bose-Einstein condensates have attracted much attention due to their promised sensitivity but are challenged by complexities and low bandwidth. However, recent experiments demonstrated [127] the versatility of these devices as well as pathways toward commercialization. Gyroscopes based on cold (not ultra-cold) atoms generated from MOTs show promise for practical use, improving on bandwidth and complexity. Maturing these devices by shrinking apparatus sizes and reducing optics complexity are in progress as these systems require multi-disciplinary advances in engineering, material science, and physics to find pathways into small, ruggedized, commercialized systems. Thermal atom beam based gyroscopes are the most developed form of atom interferometer gyroscope. These devices achieve practicality through relatively low complexity, high bandwidth, precision comparable to commercially available devices, and stability exceeding what is available in commercial, classical systems. They are currently being packaged for terrestrial and space applications and have shown promise in multiple field tests.

Finally, nuclear magnetic resonance gyroscopes utilize the stable Larmor precession rate of a nuclear spin in a constant magnetic field as a reference for determining changes in rotational orientation. The nuclear precession rate is fundamentally insensitive to the acceleration and vibration of the device's housing, and the device itself has no moving parts. The current prototype is a centimeter-scale device for commercial use and offers the promise for navigation grade performance in a small, low-power package.

## 3.1 Thermal Atom Beam Gyroscopes

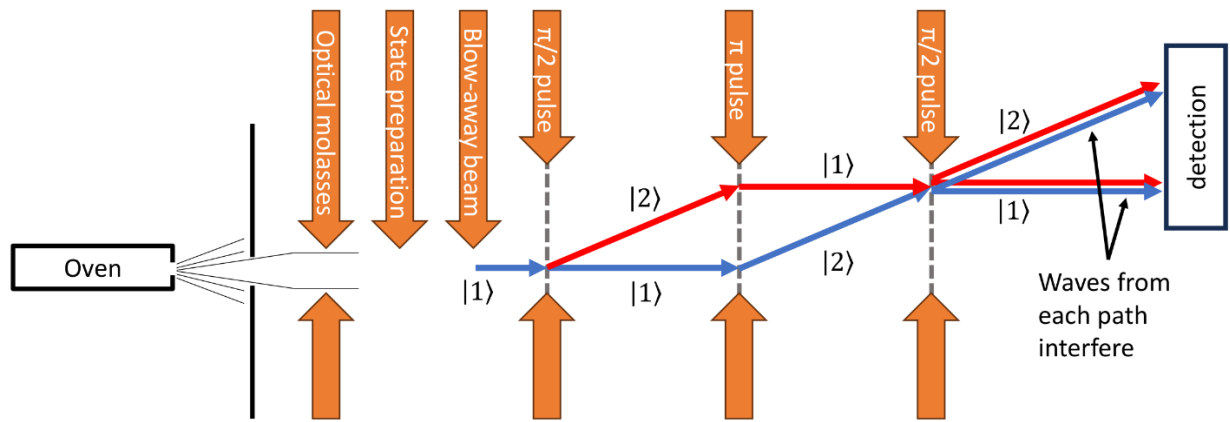

Fig. 4 Diagram of a thermal atom beam interferometer [35].

In a thermal atom beam interferometer [35], described in Fig. 4, an "oven" ejects hot atoms that pass through a collimating narrow aperture. Those atoms may optionally be cooled in transverse directions by a set of laser beams implementing optical molasses, as shown in Fig. 4, in order to further reduce transverse temperatures and increase the rate at which atoms flow through the interferometer. In general, the resulting beam of atoms with a low temperature in the vertical direction in the figure, can be passed through diffracting elements, referred to as "gratings," that split and recombine the beams to form an interference pattern. In the type of atom interferometer discussed in Fig. 4, the light-pulse atom interferometer [85, 86], those gratings are implemented by counter-propagating laser beams at different frequencies that act differently on atoms depending on what state they are in. Before the first grating, additional laser beams prepare the atoms in the beam into a desired, magnetic-field-insensitive state, and a "blow-away beam" removes atoms from the beam that failed to transfer to that state. The atom beam is then split into two different trajectories and recombined, forming a Mach-Zehnder interferometer. The numbers next to each atom beam segment indicate the internal state of atoms in that portion of the beam. The internal state of each atom is strongly correlated with its momentum (i.e. whether it is moving to the right in the figure or diagonally up and to the right). The phase of the interference pattern can be determined by using another set of one or more lasers to examine the number of atoms with each momentum or, equivalently, the number of atoms with each internal state. Thermal beam gyroscopes are furthest along in the commercialization process among atom interferometer gyroscopes. Today, the most advanced thermal beam interferometers are light pulse interferometers.

The precision of a thermal beam atom interferometer is limited by a few trade-offs centered around interrogation time and transverse temperature. In a system in which transverse laser cooling is not implemented, tightening the collimation of the beam of atoms exiting the oven can reduce the transverse temperature of the atoms. Doing so decreases statistical measurement noise associated with atoms having a diversity of trajectories creating a diversity of interference patterns. Tighter collimation can help mitigate this effect, but it will also reduce the rate at which atoms flow through the interferometer, decreasing overall signal.

Similarly, increasing the physical length of these apparatuses (to the extent that one can while still satisfying size specifications) can improve sensitivity by increasing interrogation time. However, longer interrogation times lead to other sources of signal degradation, particularly those caused by high dynamics. This degradation effect is related to both the transverse and longitudinal velocity distributions of the atoms. In addition to the effect discussed in the previous paragraph, atoms forming individual interferometers within different longitudinal velocities–and therefore different interrogation times–will experience different rotation- or acceleration-induced phase shifts. This effect similarly results in a washing-out of the interference pattern and, therefore, a reduction in precision. Some of the effects can be mitigated by using Doppler shifts to filter out atoms with velocities too far from a desired value [128, 129]. If the effects of rotations or accelerations on an atom interferometer are either applied too strongly (via high platform dynamics) or applied for too much time (via long interrogation times), the interference pattern can end up displacing so far away from the region in which it is typically measured that information is incorrect. This effect can be visualized classically in the top panel in Fig. 2 if one were to consider excessively curved classical trajectories. Thus, atom interferometers are also subject to tradeoffs between precision and dynamic range. Scientists and engineers seeking to design an inertial sensor based on thermal atom beam interferometry need to take all of these trade-offs into account along with the expected platform dynamics and find the design parameters that result in optimal performance.

Typical thermal atom beam interferometers designed to maintain a large fraction of their maximum precision at rotation rates of a few radians per second are typically able to resolve rates on the order of tens of $10^{-9}$ radians per second in 1 second of measurement. Bias stabilities on the order of tens of $10^{-9}$ radians per second have been shown, and numbers an order of magnitude better have been projected [35, 36]. The precision figure is an order of magnitude better than that of the well known GG1320AN Digital Ring Laser Gyroscope found in many aircraft IMUs [119, 120], which has similar dynamic range and bias stability. The Scalable SIRU-E has an order of magnitude better precision and bias stability than the atom interferometer but two orders of magnitude less dynamic range [125]. All aforementioned gyroscopes have scale factor stabilities on the order of a part per million. The quasi-continuous nature of the atomic interference pattern data flowing into the detector enables bandwidths in the multiple hundreds of Hz that match those of the classical sensors.

Like many cold and ultra-cold atom interferometer architectures, modern thermal beam atom interferometers serve double duty as gyroscopes and accelerometers. Rotations and accelerations both change the trajectories of atoms in an interferometer or, equivalently, shift the phases of the atom waves, resulting in a corresponding observable phase shift of the interference pattern. Therefore, it is important to be able to differentiate between the effects of rotations and accelerations on the sensor. To distinguish rotation from acceleration, a single device typically contains two counterpropagating atom beams, each being manipulated and probed by the same set of laser beams to minimize systematic errors between interferometers. Rotation of the device in the plane of the two interferometers will cause the atoms to deflect in opposite directions with respect to the apparatus, while accelerations will cause deflections in the same direction.

The high bandwidth of these systems, combined with demonstrated dynamic ranges of multiple radians per second and multiple $g$s, could serve as primary sensors in inertial navigation systems on relatively dynamic platforms, such as aircraft. There are a few companies developing thermal atom beam interferometers into small, rugged packages. Anticipation is high from the aerospace industry demanding for portable and space qualified combined gyroscope-accelerometer devices. Development efforts are targeting performance metrics that will yield a navigation error of 100 meters after 1 hour of travel in space and 30 meters after 1 hour of travel in terrestrial environments. [55]

## 3.2. Cold-Atom Gyroscopes

In cold-atom gyroscopes, MOTs are used to cool the thermal motions of atoms, reducing the spread of relative velocities in the atomic gas. For gyroscopes, this temperature reduction has several advantageous effects. Higher interrogation time is achievable, improving the resolution in rotation rate measurement. Interference patterns are less washed-out due to inhomogeneous effects of rotations and accelerations on atoms with different initial positions and velocities, improving signal-to-noise ratio. For a given interrogation time, a gyroscope's performance under dynamics is improved via a reduced velocity spread relative to thermal-atom gyroscopes [130]. The mean velocity of the atomic sample is defined by laser geometry and frequency, so that it is both accurately known and stable. Under static laboratory conditions, cold-atom gyroscopes can readily attain interrogation times greater than 100 milliseconds and thereby reach levels of sensitivity and stability that are unachievable in thermal-atom gyroscopes or only achievable in very large optical gyroscopes. Cold-atom gyroscopes may also be engineered with shorter interrogation times of a few milliseconds, appropriate for operation in relatively compact, dynamic, inertial measurement units. Implementations of cold-atom gyroscopes generally encompass one or more sources of cold atoms such as MOTs housed in ultra-high vacuum chambers [100, 131]. Atoms are collected from a vapor, cooled by combinations of laser beams and magnetic fields, and either dropped into freefall or launched on a precisely defined trajectory at relatively low velocities, typically a few meters per second (see, for example, [132]).

One benefit of doing atom interferometry with cold clouds of atoms as opposed to thermal beams is that the laser pulses can be timed in such a way that eliminates some sources of noise and long-term instability. For example, in a thermal beam atom interferometer, instability in the temperature of the oven creating the beam can cause instability in the beam velocity. The latter equates to instability in interrogation time, which in turn creates instability in the scale factor that relates the strength of an inertial force to the corresponding measured interference pattern phase shift. In contrast, cold atom clouds collected by a MOT can be launched with velocities that can be determined very accurately because they are directly related to atomic properties. Using pulsed lasers with cold atoms can also eliminate the effect discussed in the previous section in which the atoms' longitudinal velocity distribution causes loss of interference pattern visibility under high dynamics. Precisely timed laser pulses can be used to effectively eliminate atoms with longitudinal velocities too far from the average value because said atoms will not arrive in the region of the laser beam at the right time.

In cold atom and ultra-cold atom interferometers that sample forces on a shot-by-shot basis rather than measuring quasi-continuously, interrogation time leads to a trade-off between precision and bandwidth. Shorter interrogation times imply a higher sample rate, which allows the sensor to measure quantities that change more quickly and tolerate higher dynamics. On the other hand, longer interrogation times, associated with higher precision, imply a corresponding decrease in bandwidth and dynamic range and leaves the sensor more susceptible to aliasing–a type of error that occurs due to insufficient sampling rate–of noise sources and inertial signals [133]. Such aliasing can dramatically degrade gyroscope performance. Additionally, most cold-atom gyroscope architectures operate by implementing a periodic sequence of events: (1) cooling and trapping of atoms; (2) launch; (3) atomic state preparation; (4) quantum interferometry; (5) interferometer phase readout. Because the actual gyroscopic sensitivity is present only during the fourth step of the measurement sequence, the measurement suffers from "dead time," which further exacerbates issues with bandwidth and aliasing.

One way to remove the restrictions imposed by these trade-offs is to develop hybrid inertial navigation using both classical accelerometers and quantum accelerometers operating in tandem. In quantum-classical-hybrid devices, intelligent processing is employed to take advantage of the high bandwidths and dynamic ranges of classical sensors as well as the high precisions and accuracies of quantum sensors [39]. Additionally, research aimed at continuous, or zero-dead-time, operation of cold-atom gyroscopes has shown promising results and is likely to alleviate some of these concerns [134, 135]. Another way to address this problem is to use continuous beams of laser-cooled atoms [136, 137].

The atoms are prepared in a particular quantum state and then interact with the beam-splitter and mirror laser pulses characteristic of a light pulse atom interferometer described in Section 2. Finally, the interferometer phase is measured through an additional atom-laser interaction combined with light collection on a photodetector.

A handful of cold-atom gyroscope demonstrations have illustrated the technology's promise. For example, an interleaved gyroscope/accelerometer apparatus demonstrated the ability to resolve $2 \times 10^{-9}$ rad/s in 1 second of data acquisition–an order of magnitude better than typical thermal atom beam interferometers–and a bias instability of $5 \times 10^{-12}$ rad/s at a 12 Hz measurement rate without dead time [138]. It is worth noting that the improvement in precision occurred simultaneously with an improvement in dynamic range over the GG1320AN, which can only measure rotation rates as high as 2.5 Hz. This demonstration also simultaneously produced a high-precision acceleration measurement discussed in the next section. Such high-performance technology demonstrations have largely been restricted to static laboratory environments, but gyroscope architectures with smaller size and higher measurement bandwidth have made significant progress toward more practical applications [100].

Because atom interferometer gyroscopes are interferometric in nature, their output is a sinusoidal function of the rotation rate. For sufficiently large rotation rates, then, it is not possible to unambiguously determine the rotation rate input for a particular sensor output. This limits the dynamic range of the sensor. However, a number of innovative solutions to this problem have been developed over the years, including feedback control of interferometer phase, and dynamic range extension through hybridization with high-dynamic-range conventional sensors (see, e.g [7, 139, 140]).

A key advantage of cold-atom inertial sensor architectures over thermal atom beam sensors is the increase in achievable interrogation time. But, as with thermal atom beam interferometers, increasing interrogation times come at a price. In a particular dynamic environment, the practically useful value of interrogation time has an upper bound depending on sensor size and atomic temperature. For example, under 1 g of acceleration, atoms will fall by 0.5 mm in the first 10 ms and 50 mm in the first 100 ms. A hypothetical interferometer that could only tolerate 0.5 mm of displacement would therefore be limited to 1 g of dynamic range with a 10 ms interrogation time, and higher dynamic ranges could only be achieved by further reducing interrogation time. Other dynamics-related signal degradation mechanisms also limit the practically achievable interrogation time. For these reasons, cold-atom gyroscopes with long interrogation times (10's of milliseconds or more) on moving platforms are likely to require gimbal-stabilized mounting, and very long interrogation times (100's of milliseconds or more) require vibration-isolated, static operation. In general, practical cold-atom gyroscopes for dynamic applications are likely to work with interrogation times much longer than those achievable in warm atomic vapors or beams, but much shorter than those achievable in purely static applications. The high signal-to-noise ratio achievable in cold-atom gyroscopes compensates somewhat for this limitation in interrogation time.

Achieving excellent measurement stability in cold-atom gyroscopes will require reducing sensitivity to environmental parameters. Because atom interferometer gyroscopes require the atoms to traverse an extended spatial region to achieve rotational sensitivity, sensor architecture requires the relative optical path lengths of interrogating laser beams to be stable over that region as well. While the short optical path lengths of cold-atom interferometers relative to large-area optical gyroscopes reduces the challenge of stabilizing and eliminating path-length drift, building commercial sensors would require careful engineering of optical assemblies to minimize thermal variations and stress-induced deformation of optical surfaces. Certain types of environmentally-induced errors can be strongly mitigated through an error cancellation technique whereby the direction of inertial sensitivity is optoelectronically reversed periodically [141]. It may be possible to achieve unsurpassed long-term stability in cold-atom gyroscopes under laboratory conditions, but further testing and engineering will be required to achieve these results in aerospace applications.

Despite the great advantages of atomic gyroscopes employing atomic gasses with narrow velocity distributions, cold-atom and ultra-cold-atom inertial sensors must still overcome some challenges before achieving practical implementation. The most pressing challenges are related to supporting technology subsystems and responses to dynamics. Laser-cooling systems add significant size, power, cost, and complexity. Laser cooling schemes typically require several frequency-stabilized, narrow-linewidth laser beams of moderate optical power on the order of 10 - 100 mW. Because most cold-atom gyroscopes require that the atoms be launched in some direction (i.e., begin the measurement with nonzero velocity) in order to achieve rotational sensitivity, gyroscope laser system complexity is necessarily increased compared with that of cold-atom accelerometers.

In addition to laser systems, supporting subsystems include ultra-high vacuum systems, magnetic field control, fast optical shutters, and sensor timing and data acquisition [139]. The improvement in size, robustness, longevity, and integrability of all of these key subsystems are areas of active, ongoing research and engineering. These subsystems must be made robust and stable in the presence of varying temperature, pressure, humidity, and vibration. These challenges are additionally shared by the community developing high-performance cold-atom clocks, and solutions developed for one technology will likely prove applicable to the other. These issues of complexity are also somewhat mitigated by the fact that cold-atom gyroscopes typically provide high-quality accelerometer output as a byproduct of their rotation rate measurements, providing two sensors for the price of one.

While cold-atom gyroscope demonstrations have produced impressive results to date, particularly in static laboratory demonstrations, research continues toward developing architectures that are optimal for inertial measurement on moving platforms. The maturation of cold-atom gyroscopes is particularly in need of continued engineering of supporting technologies, including laser and vacuum systems, to improve size, manufacturability, and reliability.

### 3.3. Ultra-Cold Atom Gyroscopes

Some of the best rotation sensitivities could be achieved using ultra-cold atoms due to their extremely small initial velocity spread. One way to implement an area-enclosing interferometer for rotation sensing is with ultra-cold atoms confined in a trap. Either magnetic fields [105, 142] or off-resonant laser beams [143] can be used to trap atoms efficiently. In one implementation, the trap is relatively weak in one direction, and the atom waves (or, classically speaking, the atom cloud) are interfered via a typical light pulse interferometry sequence, such as that described in Section 2. To make the interferometer enclose an area to thereby measure rotation rate, the entire trapping potential is translated perpendicular to the axis of the trap while the interferometer is being implemented.

A different approach uses atoms confined in a spherically symmetric harmonic potential [127]. In such configuration, as shown in Fig. 5. the atoms are initially given a velocity kick with the same laser technique as above. The atoms slow down as they move away from the trap center, and eventually come to rest. A perpendicular laser then splits the atoms such that the two resulting packets undergo a circular orbit in the potential. After one or more complete orbits, the laser is again applied and a fraction of atoms brought back to rest. The output is detected using absorption imaging technique. Both techniques have been successfully implemented and have achieved enclosed areas of a few mm^2. Recent results using this method were able to resolve rotation rates on the order of 10^-4 rad/s after about 30 minutes of integration, and upcoming improvements are discussed in the reference [126]. A somewhat similar approach involving a ring waveguide achieved resolutions of about 10^-5 rad/s in the same amount of time and showed the ability to maintain good bias stability over a few hours [144]. Other methods for improvement have been proposed [145, 146], and it has been theorized that such a device could potentially resolve rotation rates as small as 10^-10 rad/s in a second [147].

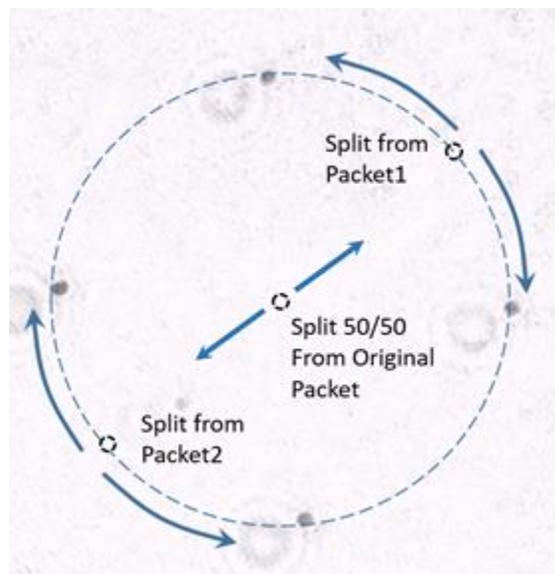

Fig. 5  Bose-Einstein Condensates (BECs) in a spherical, magnetic trap being used to implement an atom interferometer gyroscope.

Unfortunately, this performance is overshadowed by classical devices in use today. There are a few identifiable challenges to implementing a BEC atom interferometer in a portable inertial navigation system as compared to a MOT- or thermal-atom-beam-based system. The first is the production of Bose-Einstein condensates, which requires a system of lasers even more large, complex, and power-hungry than that required in a MOT-based sensor. Several lasers are required which must exhibit excellent frequency stability and tuning and whose beams must be rapidly switched on and off with good extinction. While reliable and efficient techniques have been demonstrated in laboratory settings, packaging these laser systems for use in dynamic environments will require significant engineering. In addition, the atoms must be confined in an ultra-high vacuum environment: while MOT-based systems may be able to tolerate pressures as low as $10^{-7}$ torr [100], BEC-based systems often demand background gas pressures on the order of $10^{-11}$ torr. At present there is no straightforward way to achieve this performance in a small, low-power system. However, progress on these fronts is being made, as evidenced by the recent attainment of Bose-Einstein condensation and atom interferometry in space vehicles [148] and the availability of commercial atom interferometers using laser cooled atoms.

Second, the atom trap needs to be made to be more stable as a waveguide. Magnetic atom traps, such as those in [126] are composed exclusively of magnetic fields that trap atoms in specific spin states [149]. Both electromagnetic coils and lasers used to trap atoms (such as in [143]) exhibit noise caused by power supply current fluctuations, and fluctuations in the trapping potential can lead to phase noise in the interferometer. Further, trap imperfections such as asymmetry and anharmonicity can lead to static phase shifts, and if these imperfections drift then the bias stability of the rotation sensor can degrade. Further progress will require development of both higher-stability trapping techniques and interferometer configurations which limits source of sensitivity.

Third, the sensor must be made to be more resistant to environmental effects. The quantum phase of an atom varies in response to almost any external forces, all of which must be controlled if a clean rotation signal is to be obtained. Stray magnetic fields are particularly problematic for magnetically-trapped atoms, especially over very long interrogation times (several tens to few hundreds of milliseconds) where stray fields can prevent atom wave packets from overlapping and interfering. Mitigating magnetic field noise, both from external sources as well as electronics internal to the sensor package, becomes more difficult as sensor packages become smaller, especially for ultra-cold atom systems that contain more hardware. Another challenge is dynamic range, discussed in the previous section. A potential solution is to combine the atom interferometer with a more robust but less stable sensor like an optical gyroscope. The system could be designed to rely on the optical gyroscope during periods of heavy rotations but use the atom interferometer to maintain good stability during longer periods of steady motion. There are ongoing industry efforts to demonstrate a hybrid system of this nature.

Ultra-cold atom interferometry has the potential to provide improved rotation sensors for inertial navigation, and good performance is demonstrated in a laboratory setting. A number of challenges must be resolved before the technique can be feasibly incorporated into real systems. Whether the sensor is quantum stand-alone or hybrid approach, experts from the aerospace engineering community will likely need to weigh-in for overcoming these challenges.

### 3.4. Nuclear Magnetic Resonance (NMR) Gyroscopes

An NMR gyroscope relies on measuring the precession of atomic nuclear spins about an applied magnetic field. When compared to both classical and other quantum sensors, the NMR gyroscope presents several distinct advantages, including a naturally consistent response to rotations that does not depend on the sensor size; extremely high rotation rate capability without impact on sensor noise; no moving parts; a naturally high radiation tolerance; low vibration and shock sensitivity; the capability for self-calibration; and a much simpler design than its cold-atom-based gyroscopes.

Similar to how the axis of a spinning top precesses, or rotates about the direction of gravity when its axis becomes tilted, the spin of electrons and atomic nuclei can precess at very high rates about the net magnetic field at their location. When exposed to inertial rotations about that same axis, the rate of rotation as observed in the sensor body frame appears to shift up or down proportional to the rate of rotation. While the notion of "spin" is more complicated in the quantum mechanical sense that applies to electron orbitals and atomic nuclei, the classical analog is still instructive. The core of an NMR gyroscope consists of a vapor cell filled with a gas of alkali atoms and Xenon. Xenon atoms are *spin-polarized*, or made to all spin about the same axis, via a two-step process. First, a circularly polarized light field is used to spin-polarize the electron spins of the alkali atoms. Then, through alkali-Xenon interactions, the electron spin polarization of the alkali atoms is transferred to the nuclei of the Xenon atoms. During interaction, the Xenon nuclear spin tends to align with the net magnetic field and will precess about the direction of the field until fully aligned with it. The precession phenomenon is called Larmor precession. If the apparatus is rotated, that precession frequency will change. The nuclear precession rate is fundamentally insensitive to the acceleration and vibration of the vapor cell [150]. Through the proper application of resonant transverse magnetic fields, all the polarized spins of a given Xenon isotope can be made to precess as a coherent group and provide a large signal to be read out by the alkali spins.

By using different elements and elemental isotopes in the vapor cell, the effect of environmental noise on the sensor is reduced in order to take full advantage of the stability of atomic properties [151]. The Larmor precession rate in the presence of a magnetic field is a property that can vary widely between elements and isotopes. The precession rate for the alkali atoms is roughly 1000 times higher than that of the Xenon isotopes. By monitoring how the alkali atoms' spin direction fluctuates due to perturbations in the apparent magnetic field from the precession of the Xenon isotopes, the precession frequency of each Xenon isotope can be determined. The gyromagnetic ratios for the two Xenon isotopes have opposite signs because their precessions are in opposite directions. Thanks to this physical property, if the gyroscope case is rotated at some frequency relative to a stationary inertial reference frame, the measured precession frequency for each isotope would appear to be adjusted in opposite directions by that rotation frequency when viewed from the rotating reference frame of the gyroscope case. Therefore, measurements of rotation rate are separable from magnetic field measurements and do not require high-accuracy, prior knowledge of gyromagnetic ratios. Furthermore, by reversing the relative orientation of the magnetic field and the alkali spin polarization, the major drivers of gyroscope bias instability can be reversed relative to the scale factor, making them observable and thereby reducing effective scale factor instability [152]. Utilizing this fact, bias self-calibration can be implemented on an NMR gyroscope.

Another advantage of the NMR gyroscope is that it can directly measure orientation with respect to a known starting orientation. This is an advantage over both optical and atomic interferometer gyroscopes, which only measure rotation rates and must integrate rotation rates relative to a known initial condition in order to determine their orientation at any time thereafter. The precession frequency of the Xenon isotopes is held constant by phase locking the combined signal to a stable external frequency reference and feeding back to the coils that generate the magnetic field. With the main field held constant relative to the external frequency reference, either of the two Xenon isotope signals can be used to determine the case rotation by comparing the measured Xenon isotope signal to a second stable reference signal. The relative phase angle between the measured Xenon signal and reference signal will change with a 1 to 1 correspondence as the gyroscope body is rotated. As a result, the NMR gyroscope output can be used to determine orientation angle rather than a rotation rate.

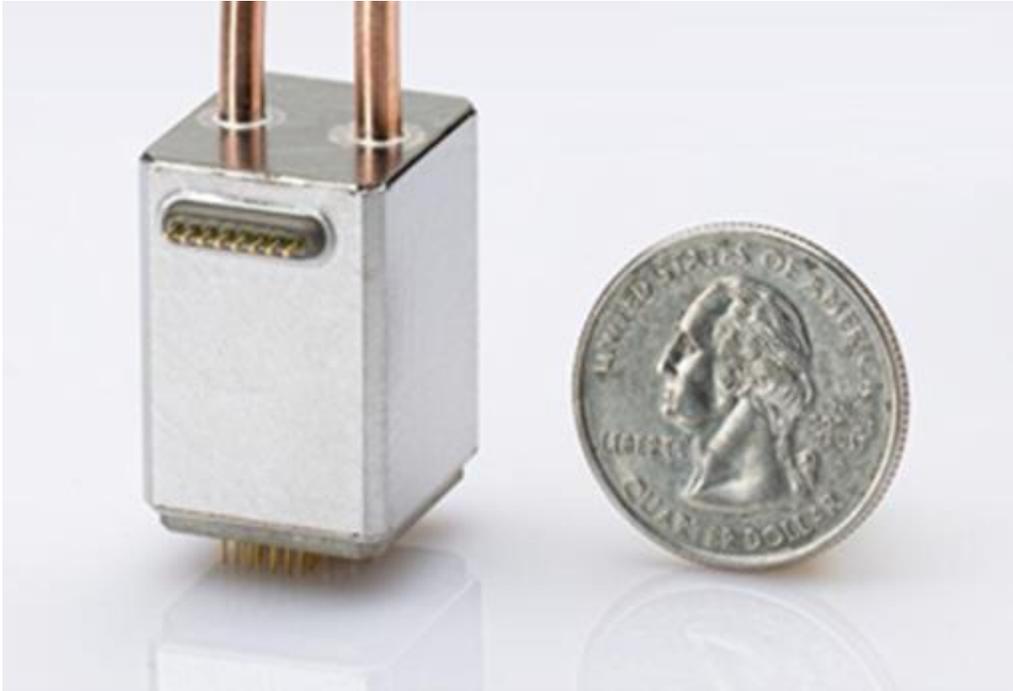

Fig. 6.  Reprint with permission from Walker, T.G., and Larsen, M.S., "Spin-Exchange-Pumped NMR Gyros," Advances in Atomic, Molecular, and Optical Physics, Vol. 65, Elsevier, 2016, pp 373-401 [150].  Photograph of NMR gyroscope developed by Northrop Grumman.

In practice, the NMR gyroscope angle measurement bandwidth is only limited by the electronics running the device, since the Xenon precession frequency can be increased simply by proportionally increasing the magnetic field. The precession frequency of the Xenon isotopes about the fixed magnetic field is independent of the motion of the detection optics or the whole of the gyroscope system around it. As a result, the changes in phase angle that are the result of any rotation of the apparatus occur instantaneously.

The scale factor, or the extent to which the device's output responds to rotations, is set by the constant properties of atoms and does not depend on any engineered properties such as enclosed area, sensitive element geometry, sensor size, or measurement time. As a result, the scale factor remains the same independent of sensor size, which is a big advantage for miniaturized sensors. This stability also supports performance beyond the state of the art even under extreme dynamics. The atoms are not aware that the sensor is rotating around them, making the NMR gyroscope capable of extreme dynamic range with maximum rotation rates that can be tens or even hundreds of Hz. This is a significant improvement over the 2.5 Hz dynamic range of the GG1320AN laser ring gyroscope. In comparison, atom interferometers often only achieve maximum tolerated rotation rates on the order of a radian per second [38, 100], though they can in principle be pushed to operate at hundreds of Hz at the cost of severe reductions in precision [153]. The atoms are constantly colliding with each other and the walls of the vapor cell, but because the nuclear spins are protected by the clouds of electrons around them, there is almost no impact on the spin coherence due to a collision. Such collisions occur roughly every 0.5 ns and have an average acceleration of greater than $10^{10}$ m/s^2, making any environmental shock or vibration, even getting shot out of a gun, a small perturbation by comparison [150].

While the reported precisions of $10^4$ rad/s of NMR gyroscopes do not exceed those of commercial devices [151], it's important to consider the extreme rotation-rate dynamic range and insensitivity to acceleration as benefits of this technology. While NMR gyroscopes may not surpass classical devices while stationary, their performance during extreme dynamics should be significantly better than atom interferometers.

With no moving parts, the primary operational lifetime limit is expected to be the lasers. Nearly all the sensor package components are naturally radiation tolerant except for the silicon photodiodes. New types of large area detectors, which are radiation hardened, are being developed to replace the silicon photodiodes and support NMR gyroscope applications in space and other high-radiation environments. The NMR gyroscope requires only one or two laser wavelengths, depending on design specifics, with modest linewidth and power requirements, and can be constructed in a much smaller and simpler package than some atom interferometry-based sensors. The cell volume for the atoms can also be much smaller (1 mm^3 volume vapor cells have been demonstrated), allowing extreme miniaturization. The application of new micro- and batch-fabrication methods to NMR gyroscope technology holds great promise for navigation using low size, weight, and power packages.

## 4. Accelerometers

An accelerometer is used to measure the change in the velocity of an object. Alongside gyroscopes, accelerometers are the other key components in inertial navigation systems. (Gravimeters, which this article classifies as a specialized type of accelerometer, may also see future use in inertial navigation systems and will be discussed in the next section). The majority of commercially available inertial measurement units are composed of micro-electromechanical system (MEMS) accelerometers [154, 155], with higher-quality devices such as pendulous integrating gyroscopic accelerometers (PIGAs) historically being reserved for certain military applications [156]. Renewed interest in long-term inertial navigation of aeronautical systems as well as inertial navigation of space vehicles requires high precision, accuracy, and stability where quantum sensors could play a vital role. The development of a high-performance accelerometer will have an impact on the next generation of navigation systems operable in space or for long durations in places where GPS is not available.  The current state of the art is summarized in Fig. 7.

Today's quantum accelerometers are based on the same set of technologies and techniques as the quantum gyroscopes described in Section 3.0, including light pulse atom interferometry. Just as small rotations of interferometer can induce changes in trajectory and, therefore, shifts in the interference pattern, small accelerations also have the same effect. For that reason, many atom interferometer inertial sensors double as gyroscopes and accelerometers. Therefore, the dramatic advancements of atom interferometer inertial sensors over the last three decades have largely benefited quantum gyroscopes, accelerometers, and gravimeters, and both are expected to develop to a new generation of devices for self-contained, inertial navigation. They also suffer from the same drawbacks, such as the trade-off between precision and dynamic range and the tradeoff between precision and bandwidth that applies to shot-by-shot, cold- and ultra-cold atom interferometers.

In steady-state laboratory conditions, atom interferometer inertial sensors can typically resolve accelerations from several hundred nano-g to a few micro-g in 1 second and can have bias stabilities on the order of micro-$g$s [35, 36, 37, 38, 157]. Compared to the Honeywell QA series accelerometers [157 - 159], which are aerospace industry staples, the thermal atom interferometers have similar precision but two orders of magnitude better bias stability. As with gyroscopes, current research focuses on making these sensors operable in dynamic environments and reducing the size, weight, power, and cost of multi-axis systems.

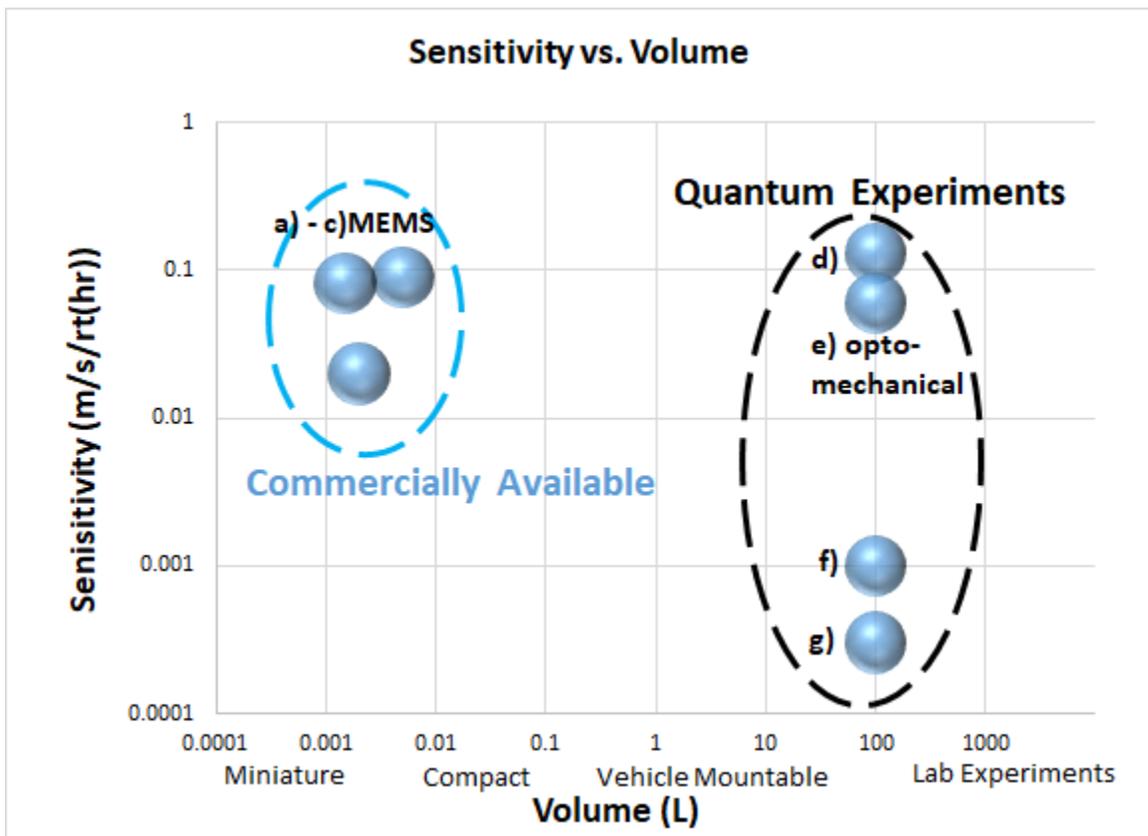

Fig. 7. Accelerometer precision, or sensitivity, vs full-system volume for both commercially available accelerometers and quantum accelerometers currently being commercialized. References: a) [160], b) [161], c) [162], d) [153], e) [163], f) [164], g) [38].

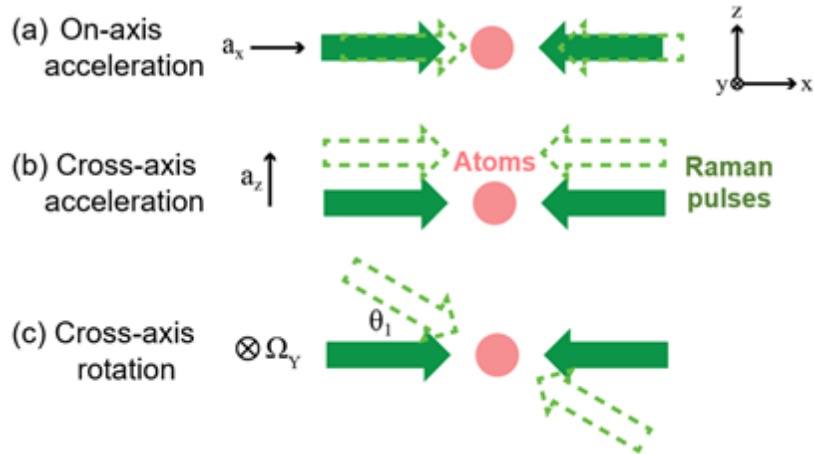

Fig. 8  The operation of a cold-atom quantum accelerometer under three representative platform movements in a dynamic environment: (a) on-axis acceleration, (b) cross-axis acceleration, and (c) cross-axis rotation.

While the trade-off between precision and dynamic range must always be recognized, there are ways to mitigate its detrimental effects. All of those effects stem from the fact that the interference signal measured by the laser system is degraded by the relative movement between atoms (released from the sensor platform and propagating freely through space within the sensor) and the light pulses (which are treated as being in a fixed position with respect to the platform). As shown in Fig. 8, there will be representative relative movements between the atoms and light resulting from the accelerations in a dynamic environment. On-axis acceleration leads to "fringe hopping," in which the system that measures and processes the phase of the atomic interference pattern loses track of how many spatial periods of oscillating bright and dark spots in the pattern have passed through the interrogating laser beams. Put simply, fringe hopping occurs when the phase of the interference pattern changes too quickly for the sensor to track those changes. Cross-axis acceleration results in lateral atomic movement, which can effectively pull the atoms to different regions in the laser beams–which have different laser intensities–that are used to measure their interference pattern. When the atoms are subjected to different laser intensities present in different regions of the beams, the resulting changes in the intended atom-light interaction create changes in the interference pattern phase that can be mistaken for signals of inertial forces. In the extreme case, excessive cross-axis acceleration can pull atoms entirely out of the laser beams that are used to measure their interference pattern. Cross-axis rotation creates spatial-overlap mismatch, which can distort and wash out the properties of the interference pattern measured by the laser system. Planar waves that would normally interfere constructively or destructively to some extent over the entire effective area of the wave instead interfere differently at different transverse locations within the wave due to cross-axis rotation. A detection system that attempts to sample the phase of the entire interference pattern will report false phases due to the presence of a continuously varying phase across the interference pattern, rather than a constant phase. Finally, mechanical vibration of the lasers and optics in the device induces noise on the interference pattern. These relative motions will all degrade the signal obtained from the atom interferometer to some degree.

It is imperative to find a mitigation strategy against acceleration- and rotation-induced degradation of the interferometer signal. A simple but powerful method to overcome these relative movements in dynamic environments is high data-rate operation using a short-time-of-flight approach. A short interrogation time between light pulses reduces the net effect of accelerations on the classical trajectories of the atom waves, curtailing significant and uncontrolled excursions of the atom clouds that result in the aforementioned degradations. Additionally, high-efficiency recapture and recycling of the cold atom clouds used in the interferometer can be used to maintain a reasonably strong signal (resulting from a reasonably large number of atoms interfered) while minimizing the dead time between measurements. However, high data-rate operation with short interrogation times leads to loss of sensitivity for the same reason: the atoms spend less time being exposed to, and therefore deflected by, the accelerations being measured. So far, high-precision, light pulse atom interferometers in packages the size of a few liters have demonstrated 50-330 Hz data rates by quickly recapturing and reusing cold atom clouds generated in MOTs [38, 100]. These atomic accelerometers show significant robustness against dynamics while still maintaining the ability to resolve micro-g accelerations in a second.

A sophisticated solution to resolve the problem of interference pattern degradation involves a feed-forward system between classical accelerometers with high data rates and a quantum accelerometer with high sensitivity. In dynamic environments, the relative movements between the atoms and light in the atom interferometer can be measured quickly (though inaccurately) with a classical accelerometer. That information can be used to apply corrections to laser frequencies and phases to maintain interference pattern visibility. Researchers have used a many-atom simulation [165] to study how dynamics affect a light pulse atom interferometer and investigate strategies to alleviate these constraints, extending the dynamic range while maintaining performance. The success of a light pulse atom interferometer will depend on mitigating the challenging dynamics through accurate modeling of the physics and implementing appropriate control algorithms through a real-time embedded processor. Dynamic ranges for on-axis acceleration can be extended by controlling the intensity and detuning of light pulses and by adjusting the velocities of atoms launched from the MOTs to reduce the on-axis relative movement. The effects of cross-axis motion can be reduced by beam steering/translating and/or gimbal stabilizing the apparatus. And the classical co-sensor can be used to measure and reject common-mode noise from mechanical vibration [164]. In particular, the feed-forward processor will recognize the motion of the sensor platform from the on-board classical sensors, estimate the physical state of the atoms within the interferometer, and execute feed-forward processing with control algorithms to provide a meaningful read out. This process will be crucial because rotations also affect the precision and accuracy of an accelerometer.

A portable, cold-atom accelerometer requires miniaturization and ruggedization of the sensor head and sub-systems. A compact sensor head of a light pulse atom interferometer [153] is implemented using a small physics package, a miniaturized, high-data-rate grating MOT, and a photonic-integrated-circuit-based laser system. Grating MOTs [166, 167] use planar diffraction grating structures, now commercially available, to create MOTs using only a single laser beam rather than two, three, or six beams in different orientations [168], greatly simplifying MOT-based systems. A passively pumped vacuum package [114] is used to validate long hold-time in conjunction with a MOT over the course of over 500 days. Chip-scale, photonic integrated circuits [110, 169] are not yet commercially viable but continue to develop at a fast pace, as problems with photonic integration at typical atomic sensing frequencies are addressed. On-chip laser system components such as single-sideband modulators, amplifiers, and nonlinear frequency doublers, have been tested against vibration, shock, and radiation [110] and used to implement atom interferometry [109]. Chip-scale atom traps, which hold atoms in optical waveguides on the surface of a photonic integrated circuit [170], are being designed and fabricated in laboratory settings and may lead to further size and power reduction of the sensor head. In the near future, chip-scale laser systems will be available for many compact atomic sensors. Today's laser systems can cost hundreds of thousands of dollars, consume multiple kilowatts of power, and take up several cubic feet of space. Advancements with photonic integrated circuits will likely lead to chip-scale, mass-produced laser systems that cost less than a thousand dollars per chip and consume under a hundred watts.

## 5. Gravimeters and Gravity Gradiometers

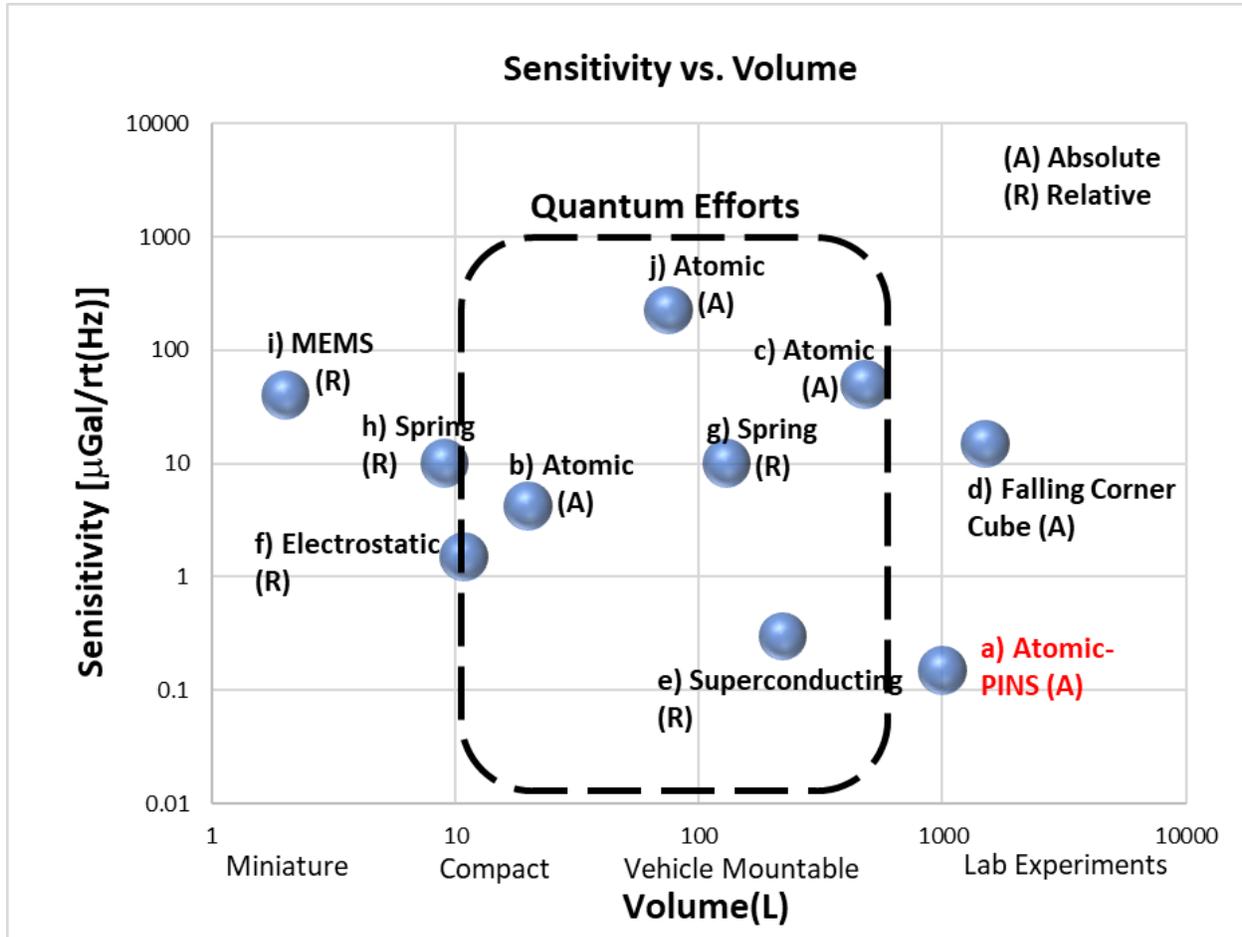

Fig. 9. Gravimeter precision, or sensitivity, vs full-system volume for classical devices and sensor-only volume for quantum devices. (A) = absolute, (R) = relative. References: a) [118], b) [171], c) [33], d) [172], e) [173], f) [174], g) [175],h) [176], i) [177], j) [51].  The dotted rectangle labeled "Quantum Efforts" also includes classical results.

Gravimeters can be thought of as specialized accelerometers that are optimized to measure Earth's gravitational pull. Gravimeters have a variety of uses in navigation, gravitational mapping, prospecting, and other applications due to their ability to measure gravity gradients, or tiny changes in gravitational acceleration over small distances. Fig. 9 shows a comparison between the state of the art of classical gravimeters and those that rely on quantum technologies.

While accelerometers and gyroscopes are core devices in inertial navigation systems, gravimeters could also play an important role in the future. During inertial navigation, it may become necessary to determine if an acceleration measured by an accelerometer is due to actual motion of the navigating craft or due to gravity pulling on the accelerometer. Unfortunately, according to the equivalence principle of general relativity [52], it is impossible for a single accelerometer to distinguish between inertial and gravitational acceleration.

A rudimentary way to solve this problem would be to have prior knowledge of the gravitational field. The known gravitational acceleration in the area could be subtracted from the measured acceleration to determine the inertial acceleration of the navigating craft, or gravimetric measurements could be compared to maps of known gravitational features in the area to determine location. However, obtaining gravity maps that are accurate enough for that application can be extremely difficult for a variety of reasons. Additionally, geological effects such as glacial movements and melting, tectonic shifts, or changes in the water table can change the local gravitational field enough to cause significant navigation errors after enough time.

Alternatively, multiple gravimeters at different locations on a craft can be used to measure gravity gradients. Accelerations of the craft will appear uniformly on all gravimeters, but gravitational accelerations from Earth or nearby objects will create gravitational fields that vary slightly across the craft. These gradients can be measured by systems of gravimeters to identify the sources of gravitational acceleration and distinguish them from inertial forces.

Gravity and gravity gradiometry also have terrestrial applications. Measuring the acceleration of the Earth's gravity is the only technique that is directly sensitive to the distribution of bulk masses surrounding the measurement platform. A gravimeter measuring the force of gravity at different locations in a small area can detect volumes within the Earth that have different densities. As a result, gravimeters provide spatial and temporal gravity data that gives access to underground density profiles and their evolution over time. Density is a very valuable observable when prospecting for underground natural resources such as water, oil, natural gas, and ore deposits, and can be used to survey for construction. The oil and gas industry is investing in commercial development of mobile gravimeters and gravity gradiometers for this purpose. Gravity gradiometry can also be used for geodesy in the measurement of the motion of tectonic plates and tides, glacial movements, as well as the polar motion of the Earth. Researchers have theorized that gravity gradiometry could also be used by customs and border protection to locate man-made tunnels or to detect hidden, high-density cargo in vehicles.

Gravimeters are traditionally divided into two complementary categories: relative and absolute. Relative gravimeters are suitable for mobile operation because they can be compact and precise, but they require initial calibration and are affected by measurement drifts. Absolute gravimeters may require little to no initial calibration and have considerably lower measurement drift over time, but are much harder to operate and tend to be less mobile. While classical gravimeters have been operated worldwide for decades, cold-atom gravimeters represent a new generation of absolute instruments that have recently left the laboratory and are seeing operation in the field [26].

High-precision, high-accuracy, absolute gravimeters rely on the same atom interferometry techniques as the gyroscopes and accelerometers discussed in previous sections. The key difference between a dedicated gravimeter and a general-purpose accelerometer is that the accelerometer is designed to function over a (hopefully) wide range of accelerations with different magnitudes and directions, whereas the gravimeter is optimized to only measure gravity that points in a specific direction with respect to the apparatus. The previous sections discuss the trade-off between acceleration and dynamic range that exists for atom interferometers and most other accelerometer architectures. Gravimeters are often designed to be about a meter long so that the atoms spend about half a second under the influence of gravity (rather than the tens or hundreds of milliseconds afforded to gyroscopes and accelerometers), leading to record-high precisions inaccessible to classical systems. As a trade-off, these devices are only capable of measuring forces that are very similar to Earth's gravitational pull, and typically only if those forces point in a very narrow range of directions with respect to the apparatus. So while their dynamic ranges are very low, these devices are capable of measuring gravity with unmatched precision and accuracy. As with gyroscopes and accelerometers, integrating quantum gravimeters with high-bandwidth, classical sensors can be used to mitigate the effects of dynamic platforms somewhat.

Measuring gravity at ground stations is the most common use case for cold-atom gravimeters today, as shown in Fig. 10. The sensor is either intended to keep measuring at one station, or it is intended to be displaced over a grid of stations–also known as surveying. Surveying can be used to prospect for oil, gas, or minerals, monitor water levels in aquifers [27], and detect subterranean voids and cavities [178]. In the terrestrial use-case, the advantages of cold-atom gravimeters come from their ability to provide precision and repeatability at a level of around $1 \times 10^{-9}$ $g$ (or one billionth of the strength of Earth's gravitational pull at sea level), even in noisy environments, which now defines the state of the art. While commercial, classical gravimeters may match that in precision [179], the key advantage of quantum gravimeters is that their drift rate can be lower than that of classical devices by multiple orders of magnitude. The main challenge in terrestrial use concerns the mobility of a sensor, which limits the number of stations at which a single sensor would be operated in a single day. While today, one can expect to make absolute measurements at around five stations per day with an atom-interferometer gravimeter, further progress needs to be realized with regard to size, weight, power, and integrability on moving platforms in order to improve their productivity.

An airborne gravimeter can provide gravity maps over large areas to benefit, for example, prospecting. Cold-atom gravimeters onboard flying platforms provide data that needs little to no calibration. The remarkable bias stability of atom-interferometer gravimeters is key to accurately imaging structures more than 10 km in size and avoiding redundant loops during the survey needed to assess the drift typical of classical gravimeters. During a first proof of concept, an absolute cold-atom gravimeter demonstrated a measurement resolution of 2 X10^-6 $g$, very close to the usual performances of relative gravimeters [28]. The static resolution of a cold-atom sensor is around 0.1 X 10^-6 $g$ typically, highlighting that, in this particular experiment, the sensor's inherent noise was not the factor limiting its resolution. Indeed, the main challenges for flying cold-atom gravimeters are the level of vibrations and the uncertainty of the knowledge of the trajectory of the carrier. Therefore, the most successful approach so far has been to fly gravity gradiometers, which rely on differential gravity measurements. Specifically, the use of a full-tensor gravity gradiometer, which provides a typical resolution of 100 X 10^-9 $g$ differences in gravitational acceleration over a few kilometers, is so far unrivaled. Measuring on-board a flying UAV is an attractive idea and can be anticipated as the maximum payload on those vehicles has increased to around 100 kg.

Gravimetry also has nautical applications. A gravimeter operated on a ship provides a map of gravity mostly defined by the seabed and the sediments below. Such gravity maps are very useful for cartography and sub-surface exploration. For nautical use, the advantage of cold-atom gravimeters is clear and has been demonstrated [29]. First, a marine cold-atom gravimeter showed superior resolution at a level of 500 X 10^-9 $g$, outperforming classical spring-based gravimeters. Second, the low bias drift of cold-atom absolute gravimeters allows operators to avoid redundant loops during the campaign needed to assess the drift of relative gravimeters, leading to shorter and cheaper surveys. A specific use case for measuring gravity at sea is navigation by matching measured gravity data with pre-existing gravity maps. In combination with inertial navigation systems used for GPS-free navigation, map-matching can bound the growth of navigational error over time that would normally exist in an inertial-only navigation system.

Operating a cold-atom gravimeter will from now on provide gravity maps with increased resolution at lower costs with the first industry-grade program recently launched in France.

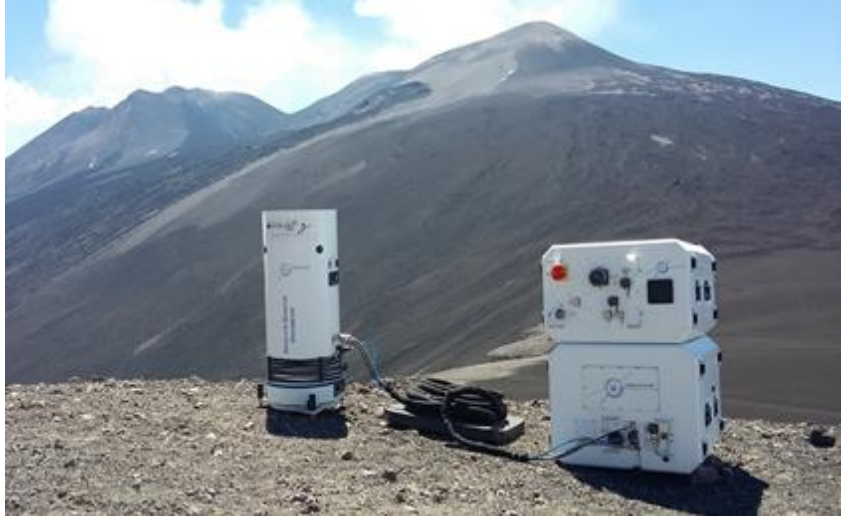

Fig. 10. Reproduced with permission from iXblue [33]. Picture of the absolute Quantum gravimeter deployed by iXblue on the volcano Mount Etna (2020).

Measuring gravity in space onboard a satellite in low Earth orbit is one of the next key use cases. Space-borne gravity gradiometer experiments are of tremendous importance to understanding the Earth and managing terrestrial resources [180]. Furthermore, now that a cold-atom experiment recently succeeded in demonstrating interferometry on board the International Space Station [148], real flying missions for the next Earth observation services are highly anticipated [181]. The community now needs to work on the size, weight, and power jointly with space agencies and satellite producers to integrate cold atom sensors into usual satellite payloads.

Measuring gravity at different length scales yields valuable information, and cold atom sensors now provide a significant technological advantage for the above use cases in real conditions. For the coming years, there is a clear need to better integrate cold-atom sensors on mobile platforms, and a better knowledge of the trajectory of the moving vehicles will be a key asset for this field.

## 6. Quantum Magnetometers

Several different types of high-precision magnetometers, including those of the quantum variety, are already commercially available, as shown in Fig. 11. Rudimentary Hall-effect types, which measure the extent to which a magnetic field changes the voltage across a conductor, are small, cheap, and easily implemented, though with less precision than their competitors. Fluxgate magnetometers [182], which measure magnetic fields by observing changes in the magnetic saturation points of certain materials, have been shown to resolve DC or quasi-DC fields such as Earth's field as small as 1 nanoTesla in only 1 second of data acquisition. SQUID (Superconducting quantum interference device) magnetometers, which measure how current flows through superconductors containing non-superconducting junctions, are widely used in medical imaging because of their high sensitivity [1, 18], though their need for cryogenic cooling necessitates bulky apparatuses. Hall and fluxgate magnetometers find uses in a widely distributed engineering community where measuring DC magnetic field down to $1 \times 10^{-9}$ Tesla is a fairly well-established practice in the presence of Earth's field, the latter of which is around 10,000 times stronger. For those needing much higher DC sensitivity ($< 10^{-9}$ Tesla) in small, low-power packages, quantum magnetometers are available.

Quantum magnetometers serve a wide variety of novel applications. Of particular interest are magnetometer systems that can be mounted on small, unmanned aerial vehicles (UAVs) for detecting obscured ferrous objects. The technique is broadly called Magnetic Anomaly Detection (MAD) and has frequently been employed for anti-submarine tactics using large aircrafts at high altitudes. With miniaturized sensors, it is possible to perform magnetic surveys from low-cost unmanned platforms.

Another application area is to use magnetometers as a navigation aid in areas where GPS is not available [183]. Scalar magnetometer readings (which describe the magnitude of a magnetic field vector) can be taken and matched with existing Earth's magnetic field map and mathematical models. Based on the comparison, a location can be determined. While magnetic navigation may underperform GPS aided inertial navigation in a short timeframe, the advantage of magnetic navigation is that the error on the resulting navigation signal stays bounded within a certain range with time. This means that magnetic navigation can serve as an important check or recalibration technique when coupled with inertial navigation systems, which normally lose accuracy at a steady rate over time (as discussed in Section 2). Magnetic navigation is enabled by quantum magnetometers by way of their high accuracy and long-term stability, which is required to make accurate comparisons to low-magnitude features on global magnetic field maps during long-duration travel, and their small size and low power requirements are necessary for integration on many aircraft.

Additional applications could involve having large arrays of networked magnetometers placed around a secure perimeter. Based on the collective readings of the magnetometers in the array, it is possible to detect a breach of the perimeter, whether above ground or subterranean [184, 185].

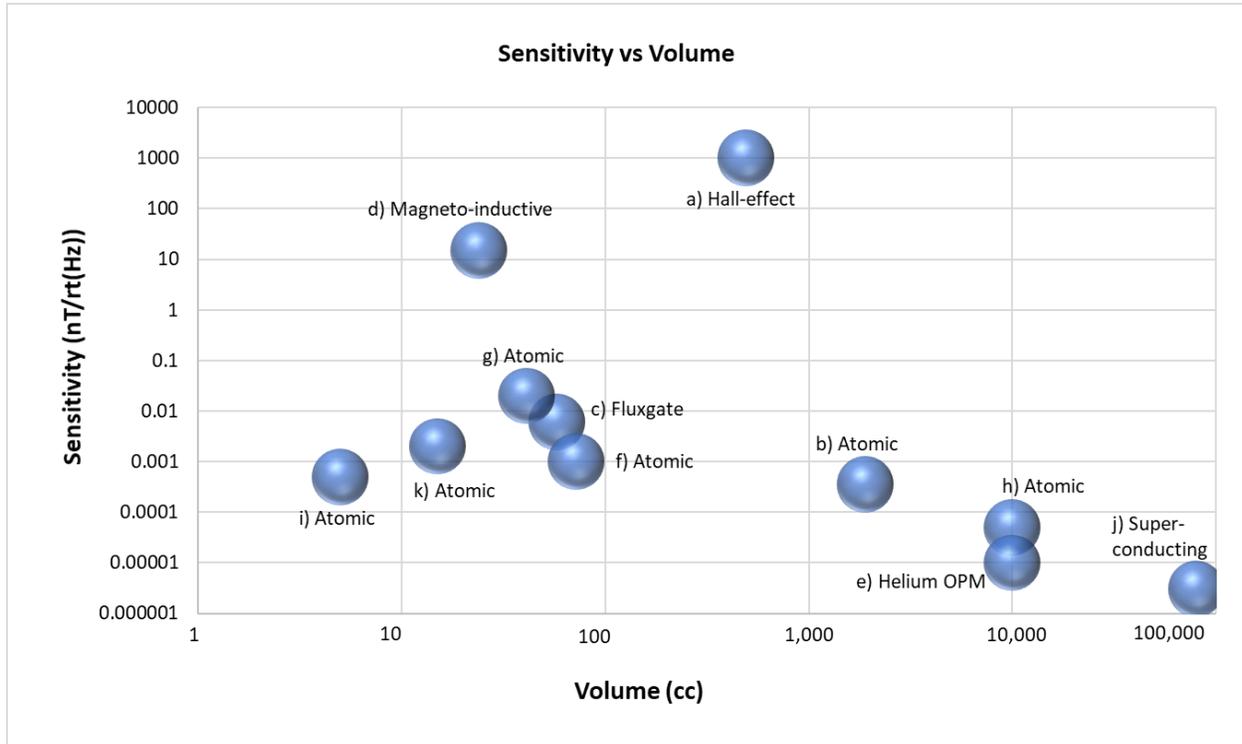

Fig. 11  Commercial magnetometer precision, or sensitivity, to DC or quasi-DC magnetic fields vs sensor volume. References: a) [186], b) [187], c) [188], d) [189], e) [190], f) [191], g) [192], h) [193], i) [21], j) [194], k) [195].

In this article, we discuss two types of quantum magnetometers that harness atomic or atom-like quantum properties. Atomic magnetometers [196 - 199] based on the vapor cells have made significant progress in the past two decades leveraging and borrowing many technologies developed from vapor-cell-based atomic clocks and are currently performing competitively with superconducting magnetometers. Another type of quantum magnetometer is spin-defect magnetometers such as those employing nitrogen-vacancy centers in diamond which are relatively new technologies that exploit the molecular structure of impurities in crystals and promise excellent precision and spatial resolution with even further-reduced complexity.

## 6.1. Atomic Vapor-Cell Magnetometers

Atomic vapor-cell magnetometers, also referred to as optically pumped magnetometers, work by using a light source to transfer a known angular momentum to atoms in a warm vapor cell [200]. The result is a collection of atoms each with a magnetic dipole moment (described in part by a spin) that, when subjected to an external magnetic field, will precess about a common axis. In a particularly simple configuration of an atomic magnetometer, as shown in Fig. 12, the light for preparing the atoms is produced from a low-power laser diode. A quarter-wave plate changes the polarization of the light from linear to circular, and the beam passes through a heated vacuum vapor cell containing an atomic gas. Due to the quantized, or discrete nature of quantum mechanics, the atoms' spins will all have the same value and will precess identically at the same frequency based on the strength of the external magnetic field. The stronger the external magnetic field, the faster the precession. The precession frequency is known as the Larmor frequency. Typically, either an applied magnetic field or the light intensity is modulated near the Larmor frequency, and when the response to the modulation is maximized, the modulation frequency is matched to the Larmor frequency, revealing the external magnetic field with high precision. Control electronics provide the signal processing to extract the magnetic field from the photodiode signal as well as temperature and frequency stabilization for the laser. This measurement technique produces precise measurements of external magnetic fields in a way that exhibits very low drift and does not require calibration. There are numerous ways to implement the atomic magnetometer with each technique seeking to optimize aspects of precision, accuracy, size, weight, and power for the targeted application [201, 202]

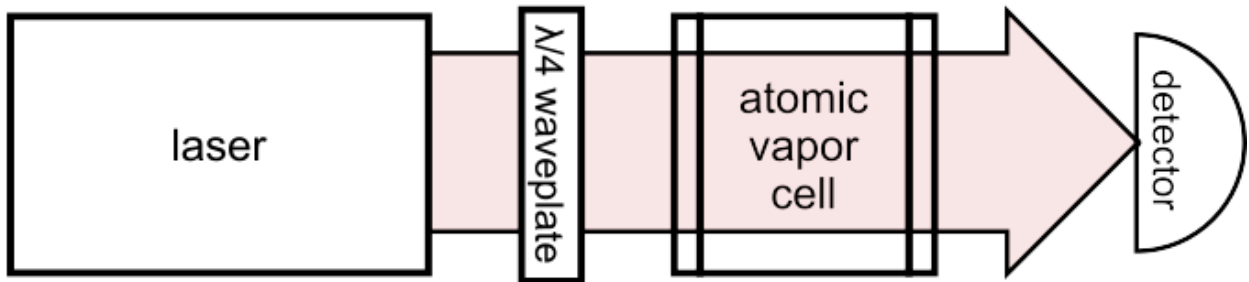

Fig. 12.  Simplified schematic of an optically-pumped magnetometer.

Traditionally, atomic magnetometers used high-power alkali lamps, large, fist-sized vapor cells, and large electronic controllers exceeding 1 liter in size, several kilograms in weight, and consuming 50 watts of power. Recent advances in chip-scale vapor cells, low-power laser light sources, and miniaturized electronics have resulted in atomic magnetometer devices that can exceed the performance of previous devices, but in a package that is significantly smaller and lighter with 50x less power consumption at a small fraction of the cost. As an example, one such device can resolve 500 femtoTesla in one second of data acquisition (500 femtoTesla is around 100 million times weaker than Earth's field) in a small package weighing only 10g (100g with control electronics) as shown in Fig. 13 [21].

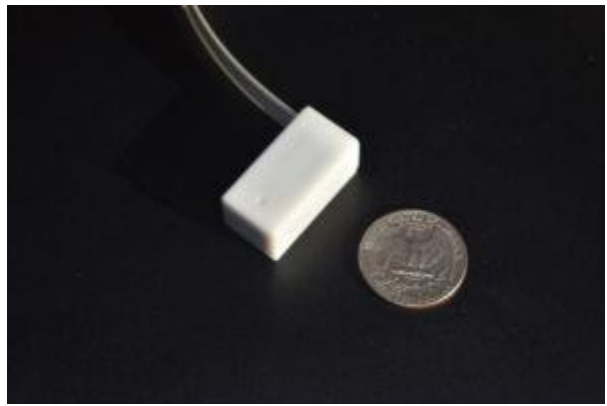

Fig. 13.  Reproduced with permission from FieldLine, Inc. [21].  An example atomic magnetometer shows how miniaturized these contemporary sensors have become.

Laboratory demonstrations of magnetometers that are less far along in the miniaturization process have achieved even better sensitivities, even going so far as being able to resolve < 10^-15 Tesla fields in a second. Initially, these magnetometers lacked the dynamic range to maintain such high sensitivities in the presence of Earth's relatively large magnetic field. The high performance of these so-called *zero-field magnetometers* would typically be showcased in a magnetically-shielded environment [15, 203 - 205].  However, more recent research has focused on *Earth's field magnetometers,* which can maintain 10^-15 Tesla sensitivities in the presence of Earth's field and other ambient magnetic field noise [206 - 208].

While the sensor technology is now advanced, there are several challenges that still remain in terms of component development.  As with atom interferometers, there is a continual need for laser sources with lower noise and increased optical power, the latter of which improves sensor performance, that can be integrated into very small sensor packages. There is also significant work that needs to be done to integrate sensors for specific applications, particularly applications that require large numbers of sensors and complex noise rejection algorithms. Lastly, the heart of the magnetometer is the vapor cell. Vapor cells are currently produced using batch fabrication methods learned from the semiconductor industry. There is still work to do to improve the processes for higher yield, lower cost, and less part-to-part variability [57].

## 6.2. Nitrogen Vacancy Diamond Magnetometers

While vapor-cell-based quantum magnetometers measure the effect of a magnetic field on the quantized energy levels of clouds of single atoms, magnetic fields can also be measured by observing their effect on the quantized energy levels of molecular structures in a crystal. More specifically, point defects in a crystal caused by the presence of impurities, also known as *color centers* or *spin-defects,* can exhibit similar optical properties as atoms and molecules, and their spin can be measured using a specific sequence of optical excitation. Many spin-defects can emit light across the visible and near-infrared spectrum, and the optical and spin properties of many of these color centers have been studied extensively [58, 209]. Diamond is an exceptional material for hosting spin defects due to its wide electronic bandgap and mechanical strength. It hosts many optically active defect centers, but the negatively charged nitrogen vacancy (NV) center stands out for its coherence at room temperature, sensitivity to magnetic fields, and ease of control and state readout. Group IV defect centers such as Silicon (Si), Germanium (Ge), Tin (Sn), and Lead (Pb) have shown great promise for quantum computing and communication technologies [210], but their short coherence times at room temperature, limits their utility as magnetometers outside of laboratory environments.

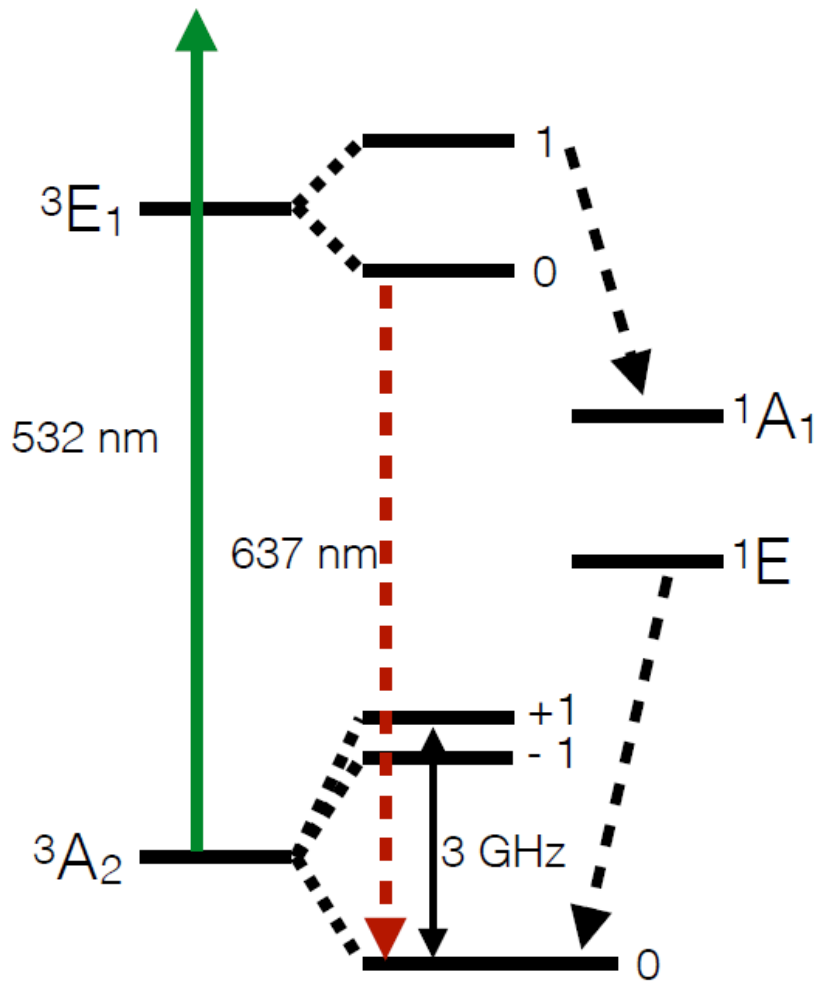

Fig. 14  Level structure of a negatively charged NV center.

An NV center in diamond consists of one carbon atom replaced by a nitrogen atom and a neighboring empty lattice site (a vacancy). These spin defects are found naturally in diamond or can be created probabilistically by nitrogen implantation and subsequent annealing. In the negative charge state, an NV center has multiple ground-state energy levels each with different spin: one with spin-0 at lower energy and two at spin -1 and +1 with higher energy, as shown in Fig. 14. An additional magnetic field will impose a relative shift in energy on the two higher-energy levels, and it is this change that is used in NV-based magnetometers [211]. When atoms are in the higher-energy ground state levels in the presence of a magnetic field, a radio frequency emitter can be used to cause the atoms to change states similar to how the quantum states of atoms in a vapor cell precess under the influence of a magnetic field. Additionally, a diamond NV center will absorb green light at 532 nm shined onto it and transition to one of the $^3E$-1 higher-energy states. That state subsequently decays through a series of non-light-emitting molecular transitions down to a lower energy excited state, which then decays back to the lowest ground state by emitting lower-energy red light at 637 nm. This green-red fluorescence is spin-dependent, implying that the strength of the red fluorescence can be used to determine which ground state the NV center is in. By shining green light on the diamond and observing red emission with a photodetector, one can measure the precession rates of NV centers in the split ground states and, therefore, the strength of the magnetic field that causes that splitting. Even at room temperature, these levels have long coherence times which can be extended by dynamical decoupling sequences, typically allowing single measurements to last as long as a few 100 microseconds [212] and even up to 1 second [213]. Thus, NV-based magnetometers provide state-of-the-art sensitivities to magnetic fields in a simple operating package consisting only of one free running diode laser, an optical detector, and a microwave generator and antenna.

NV-based sensors have demonstrated broadband detection of magnetic fields from near-DC to 10 kHz [214]. They can also be operated as magnetic field frequency detectors for signals up to the GHz range [215]. The natural distribution and orientations of NV centers has also enabled vector magnetometry [216, 217]. Bulk diamond devices can be used to address and read-out ensembles of 10,000 NV centers simultaneously, improving sensitivity as the square root of the number of NV centers increases down to a measured sensitivity of 0.3 nanoTesla/Hz^1/2 and leading to projected sensitivities of femtoTesla/Hz^1/2 [214]. In contrast, single NV centers in nanofabricated structures or nanodiamonds have been used to characterize materials and biological systems on the nanoscale [218], previously inaccessible to classical systems. Despite these impressive results, challenges still remain. NV-based magnetometers still have sensitivities orders of magnitude worse than the theoretical limits [60]. This is due to inhomogeneities in the environment across a macroscopic diamond sample, and unwanted and uncontrollable interactions between individual NV centers and between the NV centers and the surrounding environment. Work is ongoing to improve readout techniques, the quality of the diamond host material, and the coherence time of the NV centers. Moreover, while diamond magnetometers are starting to gain traction outside of the lab, more work should be done to adapt these sensors to aerospace-relevant frequency bands and to address technological limitations such as size, weight, and power.

## 7. Electrometers

Traditionally, electric field sensors, often taking the form of radio receivers, have been built using electronic circuits. A simple and common example of such is a passive circuit consisting of an antenna connected to a dissipative readout element, such as a diode. Here the propagating, free-space, oscillating, electric field that makes up a radio wave induces a potential across the antenna, which thereby collects some of the field energy. The diode rectifies the resulting alternating current signal and provides a readout voltage calibrated to convey the electric field amplitude. In fact, national standards laboratories around the world still use this technology to calibrate today's devices. Of course, such technologies have limitations. First, the conducting material making up the antenna unavoidably perturbs the field under test, making it difficult to get accurate measurements of the unperturbed field. Additionally, the size of an antenna must typically be on the order of the field's wavelength, which makes antenna construction difficult for especially long wavelengths [72]. There are challenging tradeoffs between spectral coverage (i.e., the span of usable carrier frequencies), sensitivity, instantaneous bandwidth, and dynamic range.

Individual atoms whose outer electron is highly excited, so-called *Rydberg* atoms, are now offering a radically different approach to electric field sensing. While initial research in the 1980s focused on the fundamental physics of creating highly-entangled states of light [67, 70], Rydberg-atom-based devices hold promise as electric field sensors for several reasons. Rydberg electrometers can be configured using different spectroscopy methods to operate at any carrier frequency between 0 and $10^{12}$ Hz, and they have large dynamic ranges often extending from $10^{-8}$ to $10^{3}$ V/m, giving them incredible flexibility. The individual atoms that act as electric field probes are tiny compared to the wavelength of the field under test, allowing a Rydberg electrometer configured for low-frequency use to be orders of magnitude smaller than a classical antenna. The atomic vapor and its dielectric container are essentially transparent to the fields under test. As a result, the atomic vapor does not significantly perturb the incoming electric field, which leads to the possibility of non-invasive detection of radio signals. Finally, Rydberg electrometers offer the possibility of reaching an internal noise floor below the thermal noise of its operating temperature, enabling higher sensitivity measurements. More broadly, Rydberg atoms have many appealing properties that are being investigated for a variety of quantum science applications. In addition to quantum sensing discussed here, rapid advances involving Rydberg atoms are occurring in quantum computing, information processing, simulation, quantum nonlinear optics, and transduction [68].

The outer electron of an atom can only reside in specific quantized states labeled by a *principal quantum number n* [219]. Higher energy states, with large *n*, rapidly lead to very large electron orbitals. In fact, the size of the orbital scales as $n^2$, implying that with a reasonable Rydberg state of, for example, *n* = 100, a single atom can be over a micron in size. This large separation between the valence electron and the nucleus makes these atoms highly sensitive to electric fields. Rydberg electrometers take advantage of this property to achieve high precision.

When an atom is placed in an electric field, the field will pull the negatively charged electrons and the positively charged nucleus in opposite directions, deforming the quantum states available to the electrons surrounding the nucleus that we classically think of as electron orbitals. When this occurs, the energy levels of the atom, and therefore the atom's resonant frequencies, will change. That frequency shift, known as the Stark effect, depends on the extent to which the electron orbitals are deformed by an electric field and can be described by the atom's polarizability. Due to the large separation and, therefore, weak binding between the valence electron and the nucleus, a Rydberg atom has a high polarizability and strong dipole moment, implying that its resonant frequencies will shift significantly even in the presence of a weak field [219]. By measuring those shifts in resonant frequency, the magnitude of the field that caused those shifts can be measured with very high precision.

The Stark effect can be broken down into two categories, depending if the field is near an atomic resonance ("AC") [81] or far detuned from resonance ("DC") [220]. When the AC Stark Effect is on or near resonance it is also known as the Autler-Townes effect. When exposed to an electric field oscillating on or near an atomic resonance, the energy levels associated with the resonance can "split" into multiple energy levels at slightly different energies, facilitating transitions at slightly different wavelengths. The magnitude of this resonant frequency splitting increases as $n^2$, implying that it can be quite strong for Rydberg atoms. On the other hand, when the field's frequency is far from resonance, the DC Stark Effect results in a weaker response despite its dramatic scaling of $n^7$.

A typical Rydberg sensor consists of a small glass vacuum cell ($mm^3$ to $cm^3$) containing rubidium or cesium vapor at room temperature, as shown in Fig. 15. Two semiconductor lasers together excite the atoms to Rydberg states and probe their response to an impinging external field. These sensors rely on a simple and elegant laser spectroscopy method known as Electromagnetically Induced Transparency (EIT) [221, 222]. This technique, and particularly its application in the first precision measurements of resonant radio frequency (RF) fields [223], led to a proliferation of research and attention on this new form of quantum sensor from academic groups, government labs, and private industry.

To understand EIT, consider a gas of atoms each with 3 energy levels: a ground state, a low-level excited state, and a highly excited Rydberg state. If a weak "probe" laser beam at the resonant frequency of the ground-to-low-level transition enters the atomic gas, that laser will be absorbed by the gas as it drives atomic transitions from the ground state to the low-level excited state. In a classical view, the probe laser beam transfers its energy into oscillations of atomic electrons, or *electric dipole oscillations*, and is thereby absorbed. However, if a stronger "coupling" laser beam tuned to the resonant frequency of the low-level-to-Rydberg transition is introduced, it can change the way the gas absorbs the probe beam depending on how close it is to that resonant frequency. In the presence of both probe and coupling laser beams, an additional electric dipole oscillation appears, corresponding to the low-level-to-Rydberg transition. The relative strengths of the two lasers can be chosen such that these electric dipole oscillations destructively interfere, canceling the absorption of the probe beam. In this process, the atomic population remains in the ground state and the atoms become "transparent" to the probe beam [221]. If an AC or DC electric field causes the energy level of the Rydberg state to split and/or shift, the degree to which the coupling laser beam interacts with the atoms changes, which in turn changes the amount of light absorbed by the probe beam through the EIT process.

This technique can be used to make extremely precise measurements of atomic transition frequencies. Rydberg electrometers use this technique to measure how atomic energy levels are affected by incoming electric fields, thereby measuring those fields, as shown in Fig. 16. The bottom figure in Fig. 16 shows an example of Autler-Townes splitting. Variations in the transmission of the probe beam–resonant with the ground state and principal excited state–are observed as the frequency of said beam is detuned. The blue curve is observed when no significant radio waves are present, and the green curve is observed when radio waves shift the energy level of the Rydberg state. A stronger field results in a wider splitting of the green curve as the stronger field causes a larger separation between excited-state energy sublevels in the atom.

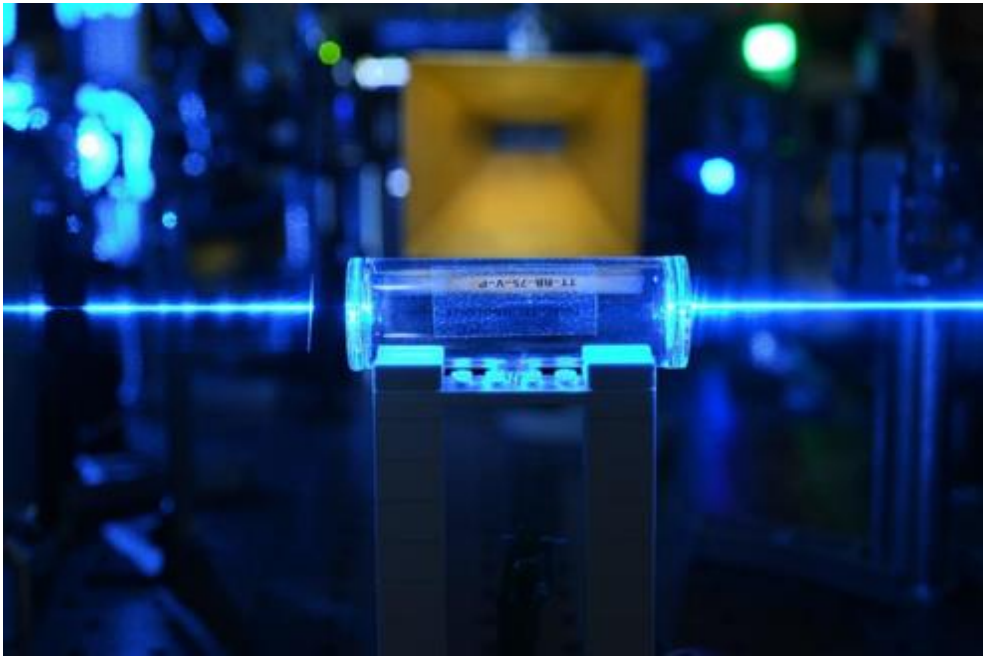

Fig. 15. Photograph by Frederik Fatemi. Photo of a Rydberg atom vapor cell. The brass-colored object in the background is a horn antenna broadcasting radio frequency (RF) signals.

Rydberg electrometers have shown significant advantages over classical antennas [226]. Reference [10] compares Rydberg electrometers with classical antennas: while classical antennas exceed Rydberg electrometers' sensitivities over a range of single frequencies, the advantage of Rydberg electrometers exists in their wide bandwidth and tuning range, low cross-section, and ability to access frequencies that are either so low that a classical antenna would be too large or so high that receiver electronics would be too difficult to construct. Precision measurements of field amplitudes become more achievable over a wide frequency range [10, 17] in addition to field imaging with sub-wavelength resolution [17]. Reception of communications data was demonstrated in 2018 [225] and achieved an instantaneous bandwidth of 5 MHz, limited by the atomic spectroscopy method. Operation over a continuous frequency span (analogous to a spectrum analyzer) with a large dynamic range was achieved by using Rydberg atoms near specially designed waveguide structures [46]. Measurements of strong fields ($5 \times 10^3$ V/m) [71], weak fields ($7.8 \times 10^{-8}$ V/m) [226], DC and low frequency fields [11], and high frequency field imaging [227] have all shown impressive performance.

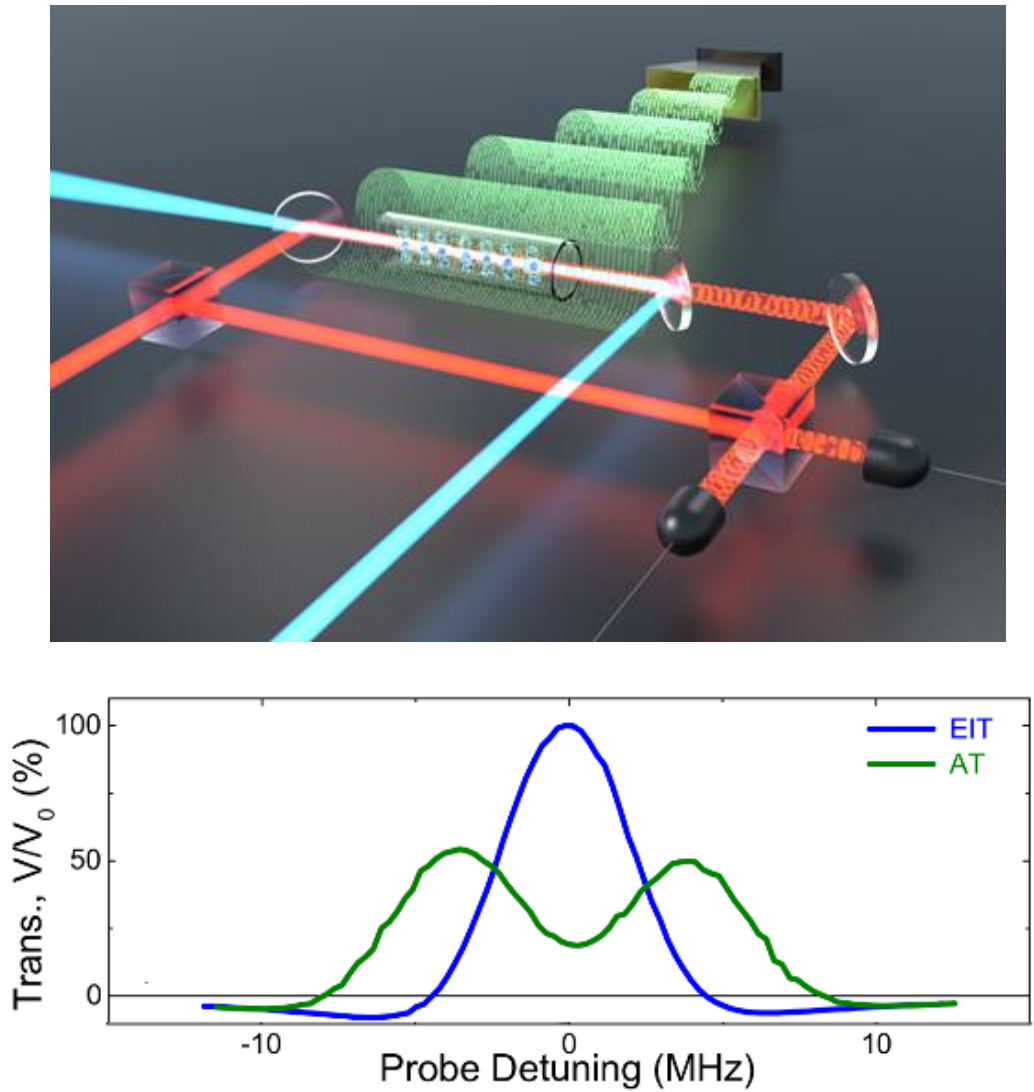

Fig. 16.  (Top) Artistic diagram of a typical Rydberg sensor experiment configuration.  (Bottom) Reprinted with permission from Meyer, D. H., Cox, K. C., Fatemi, F. K., and Kunz, P. D., "Digital Communication with Rydberg Atoms and Amplitude-Modulated Microwave Fields," Applied Physics Letters, Vol. 112, No. 21, 2018, p. 211108. https://doi.org/10.1063/1.5028357 [225].  A typical example of Autler-Townes splitting.

Progress in research and development of Rydberg sensors continues to accelerate, but this is still a young technology with many open questions and engineering challenges that must be overcome before widespread practical use becomes a reality. There is much needed exploration of potential application spaces, which in turn will inspire innovations on how to optimize Rydberg systems for specific tasks.

Because the quantum mechanics of this simple, clean system is relatively well-understood, we have a clear understanding of how close to the fundamental signal-to-noise ratio limitations we currently sit. The good news is there is still plenty of room to improve [228]. Towards the more basic-science end of the research spectrum, various approaches including different laser excitation schemes, increasing atom number, increasing coherence time, and coupling to novel waveguide and resonator structures are all being pursued.

In parallel, engineering advances of supporting technologies within Rydberg sensor systems would increase usability and encourage discovery of new applications. One advantage of Rydberg electrometers is that their sensor head, the vapor cell, can be separated from the rest of the apparatus by only an optical fiber [229]. This aspect allows for more flexible designs, including ones in which the laser and electronics are shielded from strong fields to which the sensor head is exposed and able to measure. In addition to the familiar requirements of stability, robustness, compactness, and efficiency, improvements in lasers that lead to lower noise, higher power, and greater frequency agility and modulation bandwidth would all be beneficial [10]. As an additional challenge, the atomic transition from the principle excited state to the Rydberg state often relies on laser frequencies that are much less common [230, 231] and therefore less technologically supported. However, seeing such good results in the early stages of development with such a radically different technology platform, there is much cause for optimism that Rydberg sensors will offer extraordinary value across wide areas, from communications, to biomedical imaging, to environmental monitoring, and calibration.

## 8. Outlook and Conclusion

Quantum sensing is a transformative measurement technology that promises greater accuracy, stability, and precision as well as smaller form factors all rolled into a package that is competitive with commercially available sensors. However, hype, over-promise, and reality must be examined in order to truly evaluate its full potential. While some of these sensors have realized significant commercial utility, there is still a great deal of development that is needed to mature many of these technologies to the next level of practicality. Often, we are misled with the idea that quantum sensors will replace all existing sensors; this is most likely not true. In some cases, it is likely that the unprecedented accuracy of quantum sensors will be used to supplement classical sensors, the latter of which will remain the primary workhorses. For example, optical gyroscopes and MEMS accelerometers may still be harnessed in the future for their high bandwidths, while quantum gyroscopes and accelerometers with excellent long-term stability will be used to bound the drift inherent to those classical devices.

Already, we have seen several private companies start up and begin to impact the industry. This is a testament to how these technologies will have direct benefits to aerospace and terrestrial navigation, communication, transportation, prospecting, biomedical imaging, and other sectors. The success of these efforts relies on long-term, strategic plans and investments to develop and validate supporting technologies such as compact laser systems and control electronics broadly-applicable to multiple quantum sensors. It is also important to share core technologies with other commercial platforms whenever possible, which can reduce the time and cost of rapid and agile sensor deployment for real-world applications. As much as scientists are engaged to solve technological challenges, future advances in quantum sensors likely need to be made in concert with engineers in aerospace, terrestrial, and geospatial disciplines and other fields which may benefit from these new technologies. It was mentioned multiple times throughout this article that several key developments are still needed to advance quantum technologies such as lasers, photonic integrated components, and vacuum technologies.

There are a few ways in which quantum sensor development could benefit from input from closer interaction with the aerospace industry. Because there are so many trade-offs present when designing quantum sensors–between metrics such as precision, bandwidth, dynamic range, size, power, ruggedness, and manufacturability–it is critical that the scientists and engineers designing quantum sensors engage with engineers and stakeholders in the aerospace industry to define requirements. Aerospace stakeholders can help guide the development of quantum sensing technologies by clearly defining specific applications and use-cases in which quantum sensors can provide advantages currently unavailable. Similarly, aerospace engineers can help scientists by providing specific performance and ruggedness thresholds for incorporating quantum sensors onto dynamic platforms in dynamic environments.

The aerospace industry can also assist with quantum sensor development by providing testing and evaluation opportunities at multiple stages in the development cycle. Engineering quantum sensors for realistic use cases would ideally involve frequent tests of prototypes in realistic scenarios. Lessons learned from operational tests during development could shorten the time and reduce the costs of said development.

Finally, the aerospace industry can motivate quantum sensor research by providing realistic estimates of future demand. Especially with supporting technologies such as lasers, development is hampered by a lack of investment from the private sector. Such advances require large investments and must be justified through sufficiently valuable end user application needs. Investors are hesitant to invest into technologies if there is no clear sign of high future demand. Clearly-defined demand signals from both the public and private sectors would hasten the maturation of some of these key pieces. Cognizance by the general public of quantum sensors and their benefits may also play a similar role.

One immediate application to the aerospace and transportation industry would be magnetic navigation [183], which can bring long-term accuracy to inertial navigation systems that would otherwise become useless after long-term navigation without GPS. Similarly, inertial and gravity-based navigation could also aid in high precision navigation without external aid. These technologies could one day aid self-driving cars, boats, underwater vehicles, aircraft, and even spacecraft. It is too premature to say that they could replace GPS in GPS-reliant systems, but these technologies together with inertial navigation could be a reliable alternative form of navigation when GPS is unavailable. Several demonstrations have shown the feasibility of these alternative navigation techniques in short flights through areas for which maps are available [183]. It is foreseeable that further development with the aerospace industry could improve both the sensing and the mapping capabilities required.

Recently, a collection of quantum sensors were tested at the Rim of the Pacific Naval Exercise, a multilateral collaborative exercise between the United States, Canada, United Kingdom, New Zealand, and Australia. This event allowed scientists the chance to test various portable sensor architectures in a real-life operating environment, providing critical diagnostic evaluations [232]. There are also a few efforts in the works that plan to field quantum sensors in the near future. The Defense Innovation Unit is working on building an inertial sensor based on thermal beam atom interferometry that will be qualified for use in space [55]. The plan is to fly this sensor into Earth's orbit. The National Geospatial Intelligence Agency is building a portable, high-precision gravimeter that could fit in a package the size of a 5-gallon bucket [233].

In this article, authors, experts from various quantum sensing fields, have come together to give basic understanding of the technology, its maturity, progress, challenges, as well as some possible exciting opportunities that lay ahead. We would like to take this opportunity to engage the aerospace engineering community, who may ultimately be the end users, to shape our small niche quantum community by identifying needs, specifications and performance metrics needed for practical use, and eventually transition from science to engineering. For those reasons, we hope this article will serve as a synergetic introduction to the aerospace community.

In summary, we are seeing quantum sensing technology continually evolve, have impacts in science, and find applications in commercial sectors. The hope is that, as these technologies further mature, engineers will find it easy to acquire and implement them in their respective industries.

We would like to acknowledge Grace Metcalfe and Edward Moan for their help editing this article.

Disclaimer: the views expressed are those of the authors and do not necessarily reflect the official policy or position of the Department of Defense or the U.S. Government. Further, the authors would like to clarify that sensor performance metrics may be inaccurate, such as if proprietary data or data on recent progress was not made publicly available.